\PassOptionsToPackage{draft}{hyperref}

\documentclass[fleqn,usenatbib]{mnras}

\usepackage{amsmath} 

\usepackage{newtxtext,newtxmath}

\usepackage[T1]{fontenc}
\usepackage{ae,aecompl}

\usepackage{graphicx} 
\usepackage{amssymb} 
\usepackage{multirow}
\usepackage{booktabs}
\usepackage[english]{babel}
\usepackage[export]{adjustbox}
\usepackage{subfig}
\usepackage{url}
\usepackage{hyperref}

\title[Predicting Pulsar Scintillation]{Predicting Pulsar Scintillation from
  Refractive Plasma Sheets}

\author[D. Simard \& U.-L. Pen]{
Dana Simard,$^{1,2,3}$\thanks{E-mail: simard@astro.utoronto.ca}
Ue-Li Pen$^{1,4,3,5}$
\\
$^{1}$Canadian Institute for Theoretical Astrophysics, University of
Toronto, 60 Saint George Street, Toronto, ON M5S 3H8, Canada\\
$^{2}$Department of Astronomy and Astrophysics, University of Toronto,
50 Saint George Street, Toronto, ON M5S 3H4, Canada\\
$^{3}$Dunlap Institute for Astronomy and Astrophysics, University of
Toronto, 50 Saint George Street, Toronto, ON M5S 3H4, Canada\\
$^{4}$Canadian Institute for Advanced Research, Program in Cosmology
and Gravitation, Toronto, ON M5G 1Z8, Canada\\
$^{5}$Perimeter Institute for Theoretical Physics, 31 Caroline Street North, Waterloo, ON N2L 2Y5, Canada
}

\date{Accepted XXX. Received YYY; in original form ZZZ}

\pubyear{2018}

\begin{document}
\label{firstpage}
\pagerange{\pageref{firstpage}--\pageref{lastpage}}
\maketitle

\begin{abstract}
The dynamic and secondary spectra of many pulsars show
evidence for long-lived, aligned images of the pulsar that are
stationary on a thin scattering 
sheet. 
One explanation for this phenomenon considers  
the effects of wave crests along sheets in the
ionized interstellar medium, such as those due to Alfv\'en waves
propagating along current 
sheets.  If these sheets are closely aligned to our line-of-sight to
the pulsar, high bending angles arise at the wave crests
and a selection effect causes alignment of images produced at
different crests, similar to grazing reflection off of a lake.
Using geometric optics, we develop a
simple parameterized model of these corrugated sheets that can be
constrained with a single observation and that
makes observable predictions for variations in the scintillation of the
pulsar over time and frequency.  This model reveals qualitative
differences between lensing from overdense and underdense corrugated sheets:
Only if the sheet is overdense compared to the surrounding interstellar medium can
the lensed images be brighter than the line-of-sight image to the
pulsar, and the faint lensed images are closer to the pulsar at
higher frequencies if the sheet is underdense, but at lower
frequencies if the sheet is overdense.  

\end{abstract}




\section{Introduction}
\label{sec:introduction}
Observations of pulsar scintillation, the variation in intensity over
frequency and time due to 
propagation effects induced by the interstellar medium (ISM), have
revealed significant structure in the secondary spectrum (the
2-dimensional power spectrum of the dynamic spectrum, the intensity of the
pulsar over time and frequency). In particular, a  parabolic
distribution of power in the secondary spectrum
has been found to be common in pulsars imaged with sufficient
dynamic range and resolution
\citep{putney_GBT_2005,stinebring_faint_2001}.  In some pulsars, 
inverted arclets with apexes along the main parabolic arc are also present
\citep{hill_deflection_2005,stinebring_pulsar_2007}; see
\citet[figure 1]{brisken_100_2010} for a particularly striking example.
These parabolic arcs can arise if the scattering is highly anisotropic
and localized at a thin scattering screen along our line-of-sight,
while inverted arclets in the secondary spectrum can result from
individually distinguishable images on the screen
\citep{cordes_theory_2006,walker_interpretation_2004}.  In this picture,
each inverted
arclet is due to the interference of one lensed image with the
other images of the pulsar, while the main parabola is due to the
interference of the bright, line-of-sight image of the pulsar with the
scattered images.  The discrete images have been observed to persist
for weeks \citep{hill_pulsar_2003,hill_deflection_2005} and show minute 
changes in their locations with frequency
\citep{brisken_100_2010,hill_deflection_2005}, suggesting small,
long-lived substructures within the screen, which are inconsistent
with the expected characteristics of isotropic turbulence in the ISM. 
Not only are
the substructures, observed to be $\lesssim 0.1$ AU
\citep{brisken_100_2010,hill_deflection_2005} in size, much smaller
than those expected from interstellar turbulence, but such large free
electron densities, $n_e \approx 100$ cm$^{-3}$
\citep{hill_deflection_2005}, are required to produce the observed
scattering angles that these structures would be out of pressure
equilibrium with the ISM and therefore rare, which is inconsistent with the
prevalence of pulsar scintillation arcs. 

\citet{pen_pulsar_2014} suggest that scintillation
is instead caused by corrugated sheets, such as current sheets along
which Alfv\'en waves with
amplitudes larger than the thickness (depth) of the sheet create many crests.  If the 
sheet is closely aligned with our line-of-sight to the pulsar  and corrugated 
in a perpendicular direction, a high gradient in free electron column density is achieved 
at each crest, resulting in large refraction angles near the crest. 
If many wave crests are distributed over the sheet, grazing
refraction off of the sheet will result in a linear series of 
images, in analogy to grazing reflections off waves on a lake.  (See
Fig. 1 in \cite{liu_pulsar_2016} for an example.)  This is
due to a selection effect - bending angles close to our line-of-sight
can be achieved by smaller, and more common, wave crests.  (Note that in contrast
to surface waves on a lake, in this picture the entire depth of the sheet is
perturbed by the waves, so that it resembles a flag in the wind.)  
If refraction is occurring due to a corrugated, closely-aligned thin
sheet, the curvature of the corrugations 
relieves the requirement for very high electron densities within the sheet; 
it is the combination of the
curvature and the difference in the electron density between the
sheet and the
ambient ISM that leads to large refraction angles.  This both
alleviates the tension between the prevalence of pulsar scintillation and the
large overpressures of the inferred structures in the ISM and 
has the implication that both underdense
and overdense lenses can produce the observed refraction angles; 
we will thus consider both cases here.  

A number of lensing models have already been
considered in the context of both 
pulsar scintillation and quasar extreme scattering events (ESEs)
\citep{fiedler_extreme_1987}, lensing of quasars by the ISM.
\citet{clegg_gaussian_1998} model two ESEs using a Gaussian-shaped lens
with a free
electron overdensity.  They find that the lens can produce the overall
shape of the light curves observed, but that the parameters of the
lens must be fine-tuned at each frequency band.  
\citet{pen_refractive_2012} consider the effect of a Gaussian-shaped
underdense lens on a point source such as a pulsar or quasar, and
find that a double-peaked light curve, such as those
characteristic of ESE's, is produced.
\cite{bannister_real-time_2016,tuntsov_dynamic_2015} model 
the dynamic spectrum of an ESE in an attempt to determine the electron
column density and shape of the lens.  They consider two lens shapes, one which is isotropic
and one which is anisotropic, but find that the data holds no
preference for one over the other.  They find that the
parameters used to model the ESE at one observing band are not suitable
at another band.  These results indicate that successfully modeling the lensing behaviour 
at a single frequency band and epoch is not enough to suggest consistency of the 
model with observations; a successful model must also predict changes
in the scattering with time and frequency.

In this paper, we investigate the effects of a thin, corrugated plasma
sheet closely aligned to our line-of-sight, like 
the current sheets corrugated by Alfv\'en waves 
discussed by \citet{pen_pulsar_2014}, on emission from a pulsar. 
In Section \ref{sec:lensing} we construct a
model of this lens and examine it analytically, while in Section
\ref{sec:numerical} we present some numerical examples of this model.
In both Sections \ref{sec:lensing} and \ref{sec:numerical} we examine 
the magnifications and angular 
separation between the images when multiple images form at a single
crest and we 
explore the evolution of the lensing
with time and frequency. 
Observations which can be compared to this model are discussed in
Section \ref{sec:observations} and
extensions to this model are considered in Section
\ref{sec:extensions}.  We finish with concluding remarks
in Section \ref{sec:conclusions}. 

\section{The Model}
\label{sec:lensing}

\begin{figure*}
  \centering
 \subfloat[\label{fig:multiplefolds-sub}]{\includegraphics[width=0.4\textwidth,valign=m]{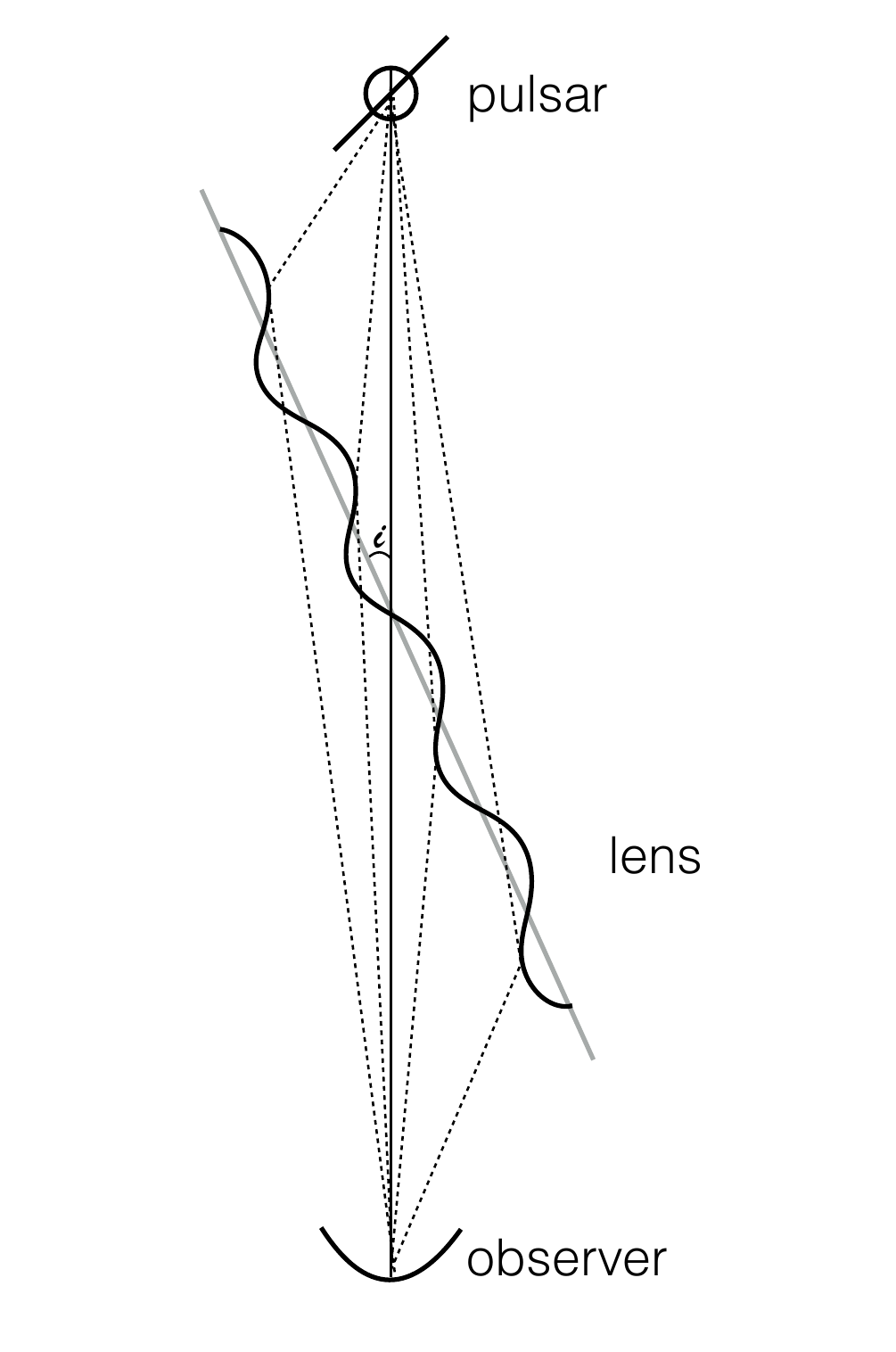}}
 \subfloat[\label{fig:roc}]{\includegraphics[width=0.5\textwidth,valign=m]{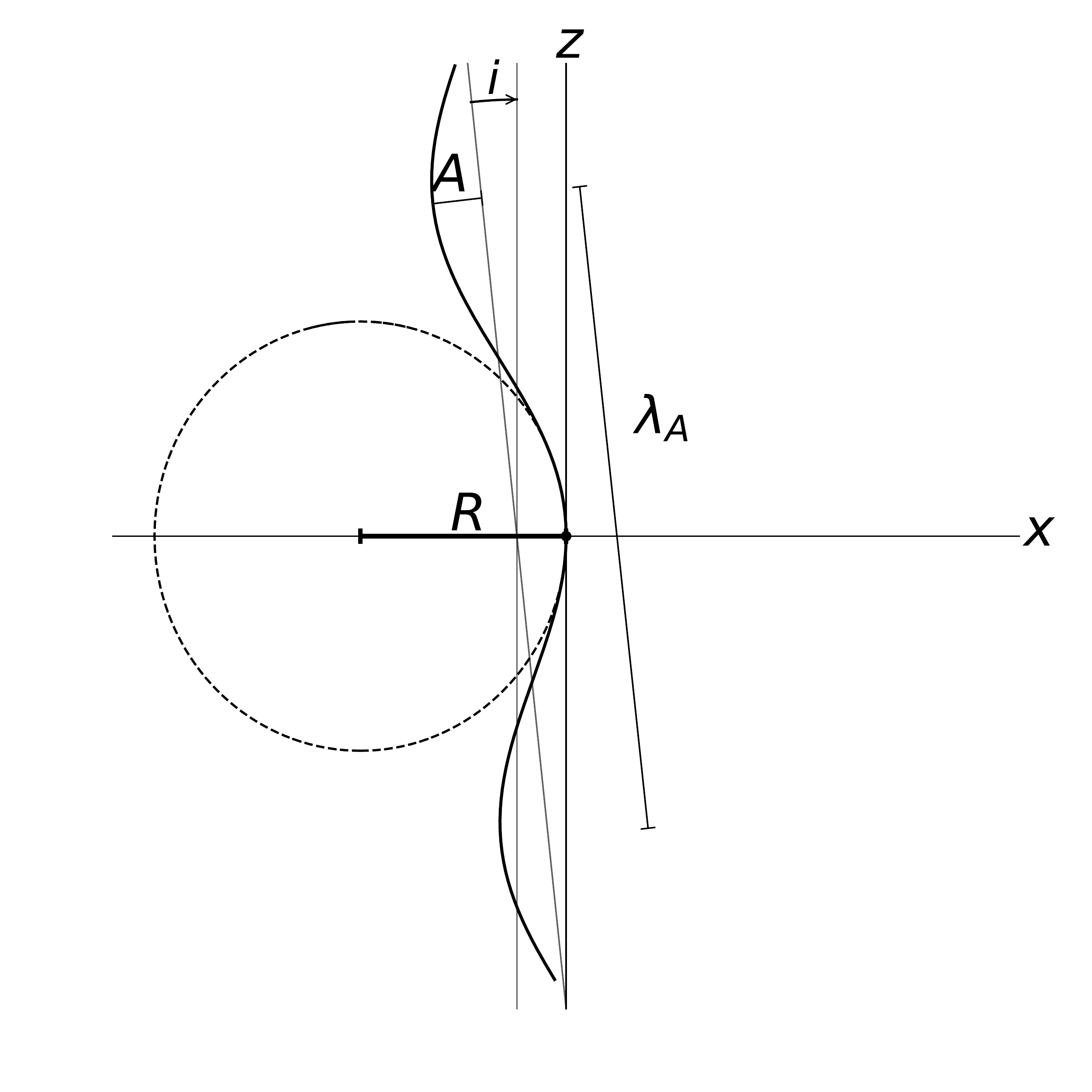}
 \vphantom{\includegraphics[width=0.4\textwidth,valign=m]{lensing_geometry.pdf}}}
  \caption{The geometry of lensing from an inclined sheet.  Figure \ref{fig:multiplefolds-sub} 
    shows how multiple wave crests result in many 
    images of the pulsar.  The line of sight image is shown with the
    solid line, while the dotted lines indicate the lensed images.  The
    grey solid line shows the orientation of the sheet, and $i$ is the
    inclination angle between the sheet and the line of sight to the
    pulsar.  The specific ray deflections shown are those that would result 
    if the corrugated sheet were underdense.
  Figure \ref{fig:roc} shows the relation between the radius of curvature at the location where the 
  column density gradient is maximized and the parameters of the wave perturbing the sheet.
   The thick curve represents the sheet, perturbed by a sinusoidal wave
 with wavelength $\lambda_A$ and amplitude $A$, and inclined by an angle $i$ relative to the
 line-of-sight to the source.  The 
 column density through the lens is maximized at the origin, and the dashed circle is
the osculating circle, with radius $R$, at this point.  Note that these figures are not to scale:  We've
drawn the lensing sheet to 
occupy the whole distance between the observer and the pulsar, while
in reality it only occupies a small percentage, we expect dozens
of crests along the current sheet rather than the few that we've
drawn, and the angle $i$ is much less than 1.  }\label{fig:multiplefolds}
\end{figure*}

In the picture of scattering from a corrugated sheet presented by
\cite{pen_pulsar_2014}, each wave crest along the inclined sheet
produces an image of the pulsar and each crest can be parameterized from properties of the wave and the observing geometry, as shown in
Fig. \ref{fig:multiplefolds}.  The lensing in this picture
can be explored in different ways.  One can
consider how properties of the sheet, including the thickness of the
sheet and the
distribution of waves, determine the statistical
properties of the scattering, such as the
angular distribution of the scattered radiation, which can be compared to
the observed scattering tails or Very Long Baseline Interferometry (VLBI) correlated flux densities of
pulsars.  In this paper we take a different approach, and focus 
on the case where individual scattered images of the pulsar are
distinguishable, for example as inverted arclets in the secondary
spectrum.  In this regime, the magnification and position of 
each image is related to properties of the crest
producing that image.
Furthermore, by tracking the image locations and magnifications through frequency and
time, the dependence of the scattering on both the observing frequency and the
proximity of the pulsar to the crest can be compared to the model.

\begin{figure*}
  \centering
 \subfloat[\label{fig:geometry_lens}]{
  \includegraphics[width=0.5\textwidth]{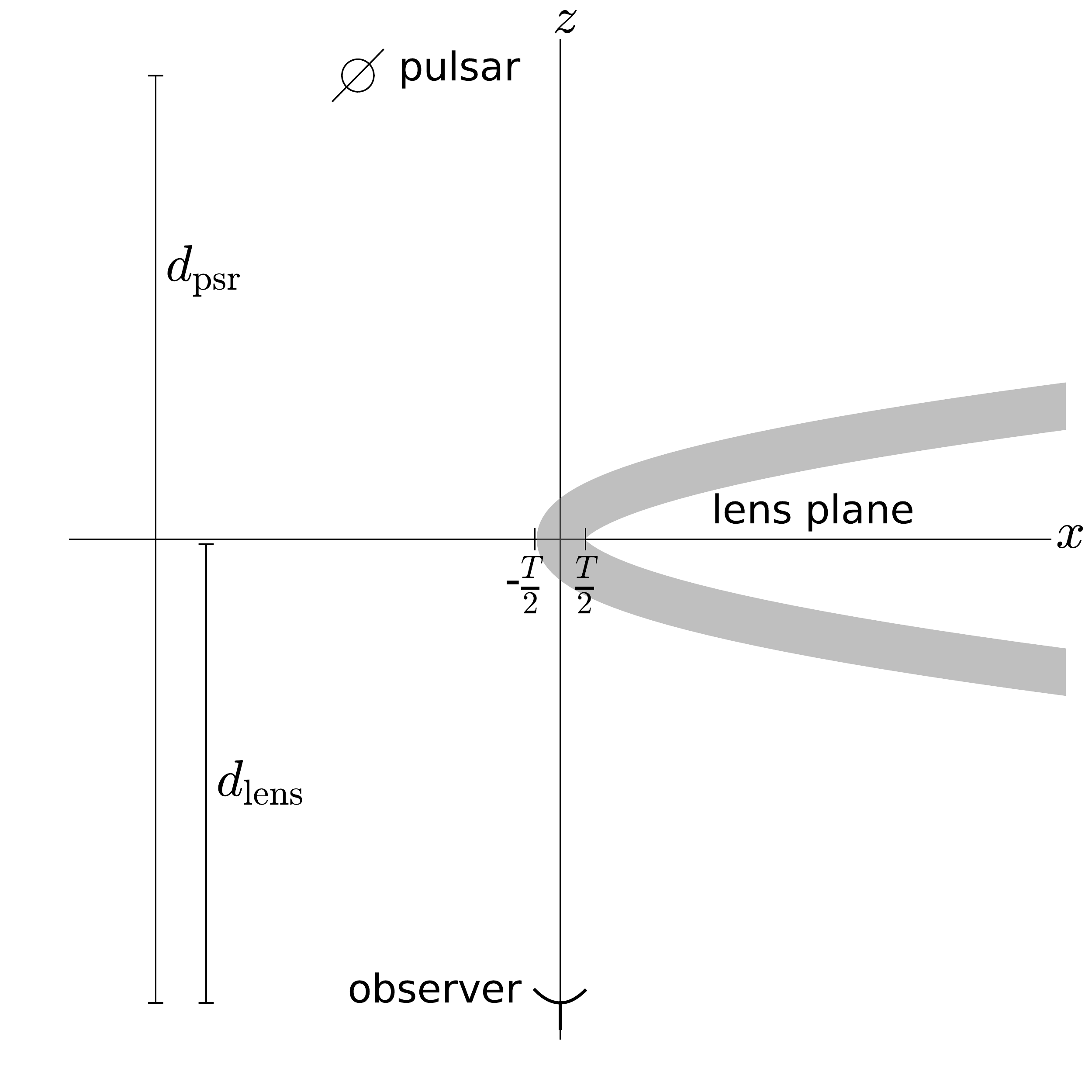}}
 \subfloat[\label{fig:geometry_density}]{
    \includegraphics[width=0.5\textwidth]{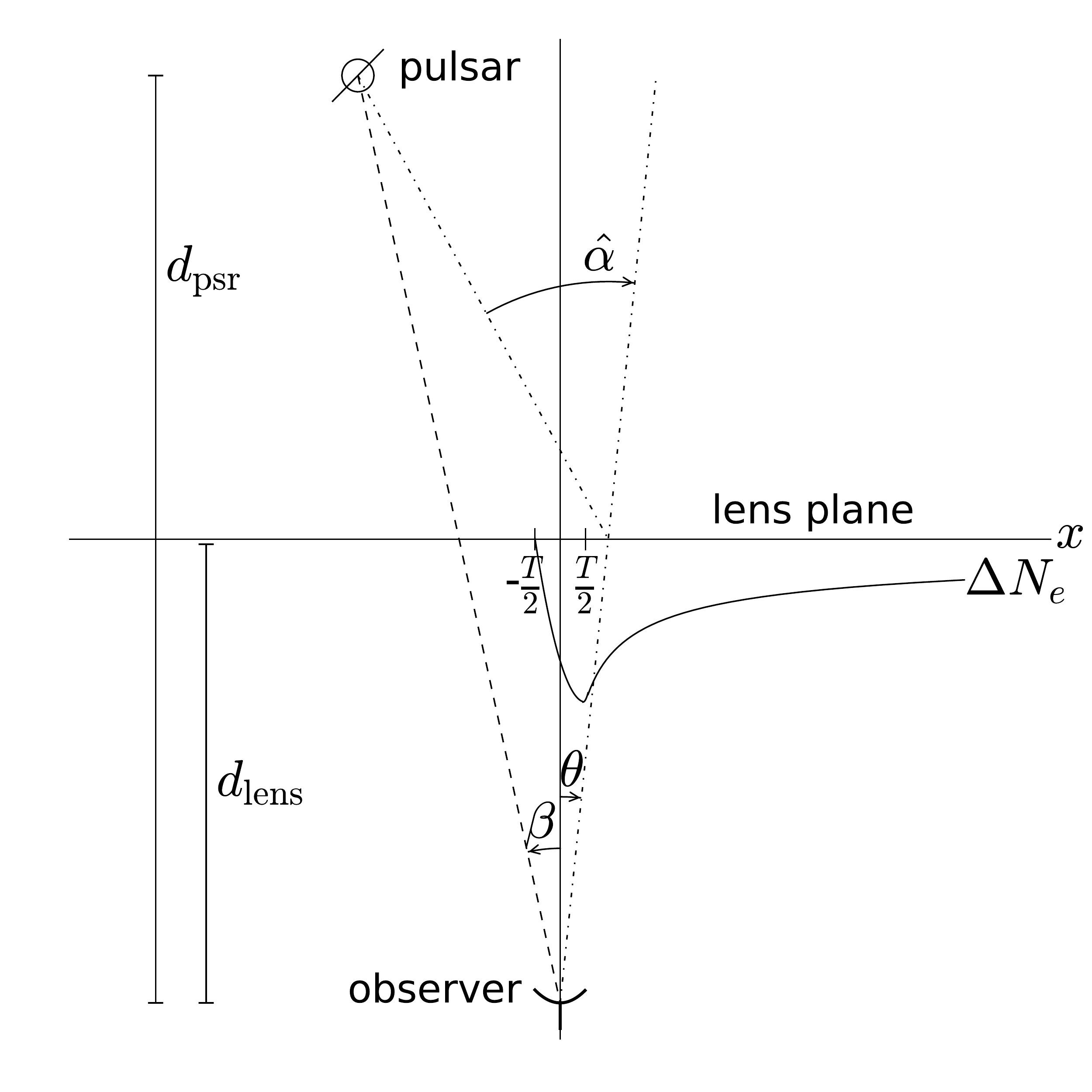}}
  \caption{ Lensing geometry and definitions of variables that we will
    use in this paper.  The grey curve in Fig. \ref{fig:geometry_lens}
    represents a single crest the sheet, while the solid curve in
    Fig. \ref{fig:geometry_density} is a sketch of the column density
    profile of the lens.  The angles $\theta$ and $\beta$ are measured
    clockwise from the line from the observer to the crest of the
    corrugation, so that $\beta<0$ as drawn.  The angle $\hat{\alpha}$ is measured clockwise from 
    the line between the pulsar and the image of the pulsar on the lens plane. 
    Note that in practice the angles are all much smaller than 1 and the crest 
    covers only a very small portion of the line-of-sight to the pulsar.   The specific deflection 
    angle and column density profile shown are those for an underdense corrugated sheet. }
\label{fig:geometry}
\end{figure*}

With this in mind, we begin our investigation
by considering a single wave crest.  We choose $z$ to be the
line-of-sight direction, and $x$ and $y$ to be in the plane of the
sky, with the origin at the crest.  We will consider a sheet of
thickness $T$ corrugated in the $x$ direction, so that all refraction
occurs in the $x$ direction.   
We will model the crest itself as a parabola in the $x$-$z$ plane.  In
truth, we do not know the 
orientation of the sheet or the corrugations in the $x$-$z$ plane, 
but since it is the projected curvature 
that determines the lensing behaviour, differences in
inclination can be accounted for by changing the curvature of the
parabola.  We will focus on a single wave crest, and write
the equation for this crest as
\begin{equation}
  \label{eqn:sheet}
  x = \frac{z^2}{2R} \;,
\end{equation}
where $R$ is the radius of curvature projected along the $x$ direction 
at the apex of the crest. 

We will use the lensing geometry shown
in Fig. \ref{fig:geometry}.  In this geometry, the lens equation is
\begin{equation}
  \label{eqn:lensing}
  \theta = \beta + s \hat{\alpha} \;,
\end{equation}
where $\theta$ is the observed position of the source, $\beta$ is the
true position, $\hat{\alpha}$ is the bending angle, $s =  1 -
d_\mathrm{lens}/d_\mathrm{psr}$, $d_\mathrm{psr}$ is the distance to 
the pulsar plane, and $d_\mathrm{lens}$ is the distance to the lens 
plane.  We determine $\hat{\alpha}$ by considering $\Phi$, the phase 
change imparted by the lens, 
\begin{equation}
  \label{eqn:phasechange}
  \Phi(x) = \frac{2 \pi}{\lambda} \int dz (n(x) - n_0) \;,
\end{equation}
where $n_0$ is the index of refraction outside of the lens, $n$ is the
index of refraction inside the lens, and $\lambda$ is the
wavelength of observations.  The bending angle is related to the
gradient of the phase change by
\begin{equation}
  \label{eqn:bending}
  \hat{\alpha}(x) = -\frac{\lambda}{2\pi} \nabla_x \Phi(x) \;.
\end{equation}
Assuming that the index of refraction is constant inside of the lens,
this reduces to
\begin{equation}
  \label{eqn:bending2}
  \hat{\alpha}(x) = - \Delta n \nabla_x Z(x) \;,
\end{equation}
where $\Delta n = n - n_0$ and $Z$ is the extent of the lens in the $z$-direction.  
In the regime $x \gg T/2$, $Z = 2 T \frac{
  \mathrm{d}l}{\mathrm{d}x}$, where  $\mathrm{d}l$ is the length
element of the 
lens, $\mathrm{d}l^2 = \mathrm{d}z^2 + \mathrm{d}x^2$.  The factor of 2 comes from
the two sides of the parabola. Using equation \eqref{eqn:sheet} for the shape of the lens, we find
\begin{equation}
  \label{eqn:thickness}
  \nabla_x Z = 2 T \frac{\mathrm{d}^2l}{\mathrm{d}x^2} = - \frac{ T R }{2 x^2 \sqrt{R/2x
      + 1}} \;,
\end{equation}
so that we can write
\begin{equation}
    \label{eqn:bending2p1}
    \hat{\alpha}(x)= \Delta n  \frac{ T R }{2 x^2 \sqrt{R/2x
      + 1}} \;.
\end{equation}

For a plasma, the index of refraction is given by
\begin{equation}
  \label{eqn:indexofrefraction}
  n = \sqrt{ 1 - \frac{\omega_p^2 }{ \omega^2}} \;,
\end{equation}
where $\omega_p = \sqrt{4 \pi c^2 r_e n_e}$ is the
characteristic frequency of the plasma expressed in terms of the
electron density, $n_e$, and the classical electron radius, $r_e
= \frac{1}{4\pi \epsilon_0}\frac{e^2}{m_e c^2}$ where $\epsilon_0$ is
the vacuum permittivity and $m_e$ is the mass of the electron.  For
$n_e = 0.03$ 
cm$^{-3}$, a typical value for the ISM, this characteristic
frequency is 1.5 kHz, much smaller than the GHz frequencies
of typical radio observations.  Therefore, we approximate the index
of refraction as
\begin{equation}
  \label{eqn:indexofrefraction3}
  n \simeq 1 - \frac{\lambda^2}{2 \pi}n_e r_e \;.
\end{equation}
We now write the bending angle in terms of the electron density,
\begin{equation}
  \label{eqn:bending4}
  \hat{\alpha}(x) = -\frac{\lambda^2}{2\pi} \Delta n_e r_e \frac{ T R }{2 x^2 \sqrt{R/2x+1}}\;,
\end{equation}
where $\Delta n_e$ is the difference between the free electron
densities inside and outside of the sheet. 
Since we measure angular positions on the sky, rather than physical 
ones, we will use the dimensionless variables $\theta =
x/d_\mathrm{lens}$, $r=R/d_\mathrm{lens}$ and
$\theta_T=T/d_\mathrm{lens}$, and write: 
\begin{equation}
  \label{eqn:bending5}
  \hat{\alpha}(\theta) = - \frac{\lambda^2}{2\pi}\Delta n_e r_e \frac{\theta_T r }{2 \theta^2}
  \frac{1}{\sqrt{r/2\theta +1}}\;.
\end{equation}
We now write the lens equation for this system,
\begin{equation}
  \label{eqn:lens}
  \theta = \beta - s \frac{\lambda^2}{2\pi}\Delta
  n_e r_e \frac{\theta_T r }{2 \theta^2}
  \frac{1}{\sqrt{r/2 \theta +1}} \;.
\end{equation}
We see that 
increasing the over or underdensity, radius
of curvature, or thickness of the lens results in a larger deflection.

Under the
approximation that $\theta \ll r/2$, equation \eqref{eqn:lens} simplifies to
\begin{align}
  \label{eqn:lens2}
  \theta &\simeq \beta - s  \frac{r_e \lambda^2}{2 \sqrt{2} \pi}
  \frac{\Delta n_e \theta_T \sqrt{r}}{\theta^{3/2}}\\
  &\simeq \beta + \frac{s}{\sqrt{2}} \frac{\Delta n \, \theta_T \sqrt{r}}{\theta^{3/2}}
  \;,
\end{align}
where we have used equation \eqref{eqn:indexofrefraction3} to write the lens equation in terms of $\Delta n$.
We see that under this approximation the lens
equation depends only on the frequency of observation, the distances to the pulsar and lens, and a
single physical parameter describing the lens itself, $\Delta n \, \theta_T
\sqrt{r}$; in other words, the column density and 
radius of curvature of the lens are degenerate.  Throughout this
paper, we will use this approximation to analyze the
behaviour of the lens; however all numerical results are calculated
using the full form of the lens equation, equation \eqref{eqn:lens}.

The solution to the lens equation gives
the observed angular positions, $\theta$, of the pulsar for a given true
angular position, $\beta$.
Due to conservation of surface brightness, the magnification of the
lensed image is
\begin{equation}
  \label{eqn:magnification}
  \mu = \bigg( \frac{d\beta}{d\theta} \bigg)^{-1} \;.
\end{equation}
Using equation \eqref{eqn:lens2}, this
evaluates to 
\begin{align}
    \label{eqn:magnification2}
    \mu &\simeq \Bigg(1+\frac{3 s}{2\sqrt{2}} \frac{ \Delta n \theta_T \sqrt{r}}{\theta^{5/2}} \Bigg)^{-1}
    \;.
\end{align}

We will now consider two possibilities - either the sheet is
underdense or it is overdense. 
We will approximate the sheet as a thin sheet to simplify the
following analytic analysis.  Within this approximation, we will
consider the lens to be at $\theta > 0$, with no lens at $\theta <
0$.  If the sheet is overdense, then the index of refraction in the sheet
is less than that in the ambient ISM ($\Delta n < 0$), and the bending angles produced
are negative (see equation \eqref{eqn:bending5}).  If instead the
sheet is underdense, $\Delta n > 0$ and the bending angles produced are positive.

We can understand the general behaviour of the lens by considering
the function $\beta(\theta)$:
\begin{equation}
  \beta = \theta - \frac{s \, \Delta n \, \theta_T \, \sqrt{r/2}}{\theta^{3/2}} \;.
\end{equation}
When $\theta$ is large,
$\beta \approx \theta$, while when $\theta$ is small, $\beta \propto
\theta^{-3/2}$ if the lens is overdense and
$\beta \propto -\, \theta^{-3/2}$ if the lens is underdense.  
There is a lensed image at $\beta = 0$ when
\begin{equation}
\theta_0 \equiv \Big(  s \, \Delta n \, \theta_T \, \sqrt{r/2}\Big)^{2/5}\;, 
\label{eqn:theta0}
\end{equation}
which has a single real
solution if the sheet is underdense but no solutions if the sheet is
overdense.  Finally, we see that local extrema occur when
$\frac{\mathrm{d}\beta}{\mathrm{d}\theta} = 0$, or 
\begin{equation}
\theta_\mathrm{ext} \equiv \Big(-\frac{3}{2} s \, \Delta n \, \theta_T \,
\sqrt{r/2}\Big)^{2/5}\;.
\end{equation}
If the lens is underdense, there are
therefore no local extrema, but if the lens is overdense there is one,
a local minimum.

Combining this information, we see that if the lens is underdense there is a
single lensed image for all $\beta$, or all true positions of the
pulsar.  In addition, the unlensed image of the pulsar is visible if
$\beta < 0$.  As a result, two images are visible if $\beta < 0$,
the unlensed image and a lensed image.  If $\beta > 0$, only one,
lensed image is seen, but it tends towards the line-of-sight position
for large $\beta$.  If the lens is overdense, then there are two
lensed images for $\beta > \beta(\theta_\mathrm{ext}) > 0$, one at
small $\theta$ and one at $\theta \approx \beta$.  There is only the
line-of-sight image at $\beta < 0$, and there are no images at all for
for $0 < \beta < \beta(\theta_\mathrm{ext})$.  

From equation \eqref{eqn:magnification2}, differences in the
brightnesses of the lensed images between the two lenses are apparent.
If the lens is underdense,
$\Delta n > 0$, and $\mu < 1$, so that lensed images brighter than the
unlensed image of the pulsar cannot be produced.  When
$\beta = 0$ and $\theta = \theta_0$, we find
that the magnification of the image is $\frac{2}{5}$,
regardless of the lens parameters.   If the
lens is overdense, $\Delta n < 0$, so that $\mu > 1$ or $\mu < 0 $.
In this case, either the lensed image is brighter than the unlensed
image of the pulsar, or the image is inverted. 

In practice, as discussed later in Section \ref{sec:observations}, we do not
measure the absolute magnifications and 
locations of the lensed images, but rather the magnification ratio and
angular displacement
between two images of the pulsar.  
To compare with
observations, we will therefore investigate the magnification ratios
and angular separations between images in the cases where two images
are formed at a single crest. 

We can derive a relation between the magnification ratio and angular
separation by assuming that the brightest image is unlensed, which for
the underdense lens is true, and for the overdense lens is a good
approximation when $ \beta \gg \big(s |\Delta n|\, \theta_T
\sqrt{r/2}\big)^{2/3}$.
We will write the location and magnification of the
brightest image as $\theta_0 \simeq \beta$ and $\mu_0
\simeq 1$ respectively.  
In this case, the angular separation is 
\begin{align} 
    \Delta \theta &\simeq \theta - \beta \simeq  \frac{s}{\sqrt{2}}\frac{\Delta n \theta_T \sqrt{r}}{\theta^{3/2}} \\
    &\simeq 3.6 \mathrm{mas} \frac{s}{1/2} \frac{\Delta n}{1.3 \mathrm{x} 10^{-11}}\frac{\theta_T}{4 \mathrm{x} 10^{-10}}\sqrt{\frac{r}{10}}\bigg( \frac{\theta}{1\mathrm{mas}} \bigg)^{-3/2} \;,
\end{align}
where $\theta$ is the angular location of the fainter image.  (See Section \ref{sec:parameters} for the origin of the fiducial values chosen for $\Delta n$, $\theta_T$, $r$ and $\theta$.  For most pulsar scintillation systems, the screen is not associated with the pulsar and is midway between the observer and the pulsar, so we choose a fiducial value $s=1/2$.)  This allows us to 
write $\theta$ in terms of
the angular separation, $\Delta \theta$:
\begin{equation}
  \label{eqn:separation2}
  \theta \simeq \Bigg(  \frac{ s}{\sqrt{2}}
  \frac{\Delta n \theta_T \sqrt{r}}{\Delta\theta} \Bigg)^{2/3} \;.
\end{equation}
In this regime, we expect 
the
magnifications of the faint images to be much less than one.  We can
therefore simplify equation \eqref{eqn:magnification2}:
\begin{align}
    \label{eqn:magnification3}
    \mu &\simeq \frac{2 \sqrt{2}}{3}\frac{1}{s} \frac{\theta^{5/2} }{\Delta n \, \theta_T \sqrt{r}}\,.
\end{align}
We now combine equations \eqref{eqn:separation2} and
\eqref{eqn:magnification3} to write the magnification ratio in terms of the
angular separation:
\begin{align}
  \label{eqn:magnification4}
  \bigg|\frac{\mu}{\mu_0}\bigg| &\simeq \frac{2}{3} \Bigg( \frac{s}{\sqrt{2}} \,
  | \Delta n|\, \theta_T \sqrt{r} \Bigg)^{2/3} |\Delta
  \theta|^{-5/3} \\
  &\simeq 0.03 \Bigg(\frac{s}{1/2}\frac{|\Delta n|}{1.3 \mathrm{x} 10^{-11} } \frac{\theta_T}{4 \mathrm{x} 10^{-10}} \sqrt{\frac{r}{10}} \Bigg)^{2/3} \Bigg| \frac{\Delta \theta}{10 \mathrm{mas}} \Bigg|^{-5/3} \;.
\end{align}
Equation \eqref{eqn:magnification4} can be rewritten as
\begin{equation}
    \label{eqn:magnification5}
 \bigg| \frac{\mu}{\mu_0} \bigg| \simeq \frac{2}{3} \bigg | \frac{ \Delta \theta_{\mathrm{ref}}}{\Delta \theta}\bigg|^{5/3} \; , 
\end{equation}
 where 
 \begin{align}
    \label{eqn:dthetaref}
        \Delta \theta_{\mathrm{ref}} &= \bigg( \frac{s}{\sqrt{2}}\, | \Delta n | \, \theta_T \sqrt{r} \bigg)^{2/5} \\
        &\simeq 1.7 \,\mathrm{mas} \,\Bigg( \frac{s}{1/2} \frac{| \Delta n |}{1.3 \mathrm{x} 10^{-11}} \frac{ \theta_T}{4 \mathrm{x} 10^{-10}} \sqrt{\frac{r}{10}}\Bigg)^{2/5}
        \;.
 \end{align}
 Therefore, we expect $\big|\frac{\mu}{\mu_0}\big| \propto |\Delta \theta|^{-5/3}$ for large
 separations.  This relation allows us to determine $\Delta
 \theta_\mathrm{ref}$, and therefore $|\Delta n| \,
 \theta_T \sqrt{r}$, from
 observations of the angular separations and relative magnifications
 of the images from a single wave crest.  Note that if the sheet is underdense, $\Delta \theta_\mathrm{ref}$ is the angular separation when the pulsar is directly behind the crest, $\Delta\theta(\beta=0) =\theta_0$.  (See equation \eqref{eqn:theta0}.)

We can also consider the frequency evolution of the
lensing qualitatively.  
The bending angle $\hat{\alpha} \propto -\frac{\Delta n }{ \theta^{3/2}}$ when $\theta \ll r/2 $  and the index of refraction scales with 
wavelength, $\Delta n \propto \lambda^2$, so that at higher frequencies, images near the crest of the wave must form
at smaller $\theta$.
In the underdense case, the
line-of-sight image of the pulsar 
is at $\beta < 0$ when a second lensed image is visible, so that as
the lensed image moves towards the crest of the wave the angular
separation between the line-of-sight and lensed images decreases.  In
contrast, if the lens is overdense two images are visible when $\beta > 0$, 
the bright image of the pulsar at approximately 
the line-of-sight to the pulsar and a faint image close to the crest of the
wave.  In this case, as the wavelength increases 
the fainter image moves to smaller $\theta$ while the brighter image moves very 
little with wavelength, so that the angular
separation between the two images increases.  Thus, we see that if the
lens is underdense, the angular separation between the two images
decreases with frequency, while if it is overdense the angular
separation increases with frequency.  This behaviour allows one to
distinguish between the two cases, and determine the sign of the
lensing parameter $\Delta n \, \theta_T \sqrt{r}$.  

\section{Numerical examples}
\label{sec:numerical}

In order to construct some numerical examples of this model, we 
first put some values to parameters in the model in Section
\ref{sec:parameters}.  We then consider the behaviour of a single wave
crest for lenses of varying strengths in Section \ref{sec:results}.
Finally, we present an example of constraining the lens parameter
and predicting the evolution of the lensed images with time and
frequency in Section \ref{sec:evolution}. 

\subsection{Numerical Parameters}
\label{sec:parameters}

For our numerical model, we will adopt parameters from one of the most
thoroughly studied examples of pulsar scintillation arcs, PSR B0834+06.  We
will assume the distances 
to the pulsar and lens are those
measured by \citet{liu_pulsar_2016} for PSR B0834+06 and one of its
lensing screens, $d_\mathrm{psr}=620$ pc, $d_\mathrm{lens} = 389$ pc.
We will assume an observing frequency of
314.5 MHz.  For a typical lensed image, we will consider an angular
displacement from the line-of-sight image of the pulsar of 10 mas and
a magnification of 0.01, parameters similar to those measured by
\citet{brisken_100_2010} for PSR B0834+06.  Finally, in order to
predict the temporal evolution of this system, we will need to know
the speed with which the pulsar moves behind the lens in the
direction of scattering.  For this, we will also use the value
measured by \citet{liu_pulsar_2016}, 172.4 km s$^{-1}$, or an angular
relative velocity of 1.12 mas/week.

We will assume values for $T$, the thickness of the sheet, and
$R$, the radius of curvature of the crest, and vary only the electron
density, and thus the index of refraction, 
inside the lens, but keep in mind that this 
model only constrains the combination of parameters
$\Delta n\, \theta_T \sqrt{r}$ and not $\Delta n$ itself. 
In physical parameters, for a sheet with an inclination angle $i$ relative to
the line-of-sight to the pulsar, the radius of curvature at the point where the 
tangent is in the $z$-direction is
\begin{align}
  \label{eqn:radius}
  R &= \frac{\lambda_A^2 }{4 \pi^2 A \cos^3(i) } \frac{1}{\sqrt{ 1 -
    \frac{\lambda_A^2}{4 \pi^2 A^2 } \tan^2(i) } }\\
      &\approx \frac{\lambda_A^2}{4 \pi^2 A} \; ,
\end{align}
where $\lambda_A$ and $A$ are respectively the wavelength and amplitude of the Alfv\'en
wave.  See Figure \ref{fig:roc} for a sketch of the wave and the resulting radius of curvature.
In order to corrugate the current sheet, the amplitude of the 
wave must be much greater than the thickness of the sheet, 
and we expect a very small
inclination angle of the sheet with respect to our line-of-sight,
$i \ll 1$, in order to produce the linear series of images observed. 
We expect the projected wavelength, $\lambda_A \sin(i)$, to be similar 
to the separation between images, which
in some cases is as small as 
0.05 AU for the PSR B0834+06 system
\citep{brisken_100_2010}. 

We choose an
Alfv\'en wavelength $\lambda_A = 10^5$ AU, an inclination of the sheet
with respect to our
line-of-sight to the pulsar $i = 10^{-5}$ rad, a thickness of the sheet 
$T = 0.03$ AU or $\theta_T = 0.08$ mas, and an amplitude of the Alfv\'en wave
$A = 0.3$ AU.  These parameters give
a projected wavelength of 1 AU, and a projected radius of curvature at the 
apex of the parabola of $R = 4.8$ kpc or $r = 12$.  
Assuming a volume filling factor of 10\% \citep{draine_physics_2011}
for the warm ionized interstellar medium, and an average electron density
throughout the entire ISM of 0.03 cm$^{-3}$, we assume a typical electron
density of 0.3 cm$^{-3}$ in the warm ionized interstellar medium.  While
  our analysis does not depend on this value, it does inform the
  physically allowed electron density differences between the lens and
  the surrounding medium:  We will 
consider electron density differences between inside and outside the lens of 
$\Delta n_e = \pm 0.3, \pm 0.003$ cm$^{-3}$.  

\subsection{Lensing from a single crest}
\label{sec:results}

\begin{figure*}
  \centering
  \subfloat[Underdense lens \label{fig:lens_convergent}]{
  \includegraphics[width=0.5\textwidth]{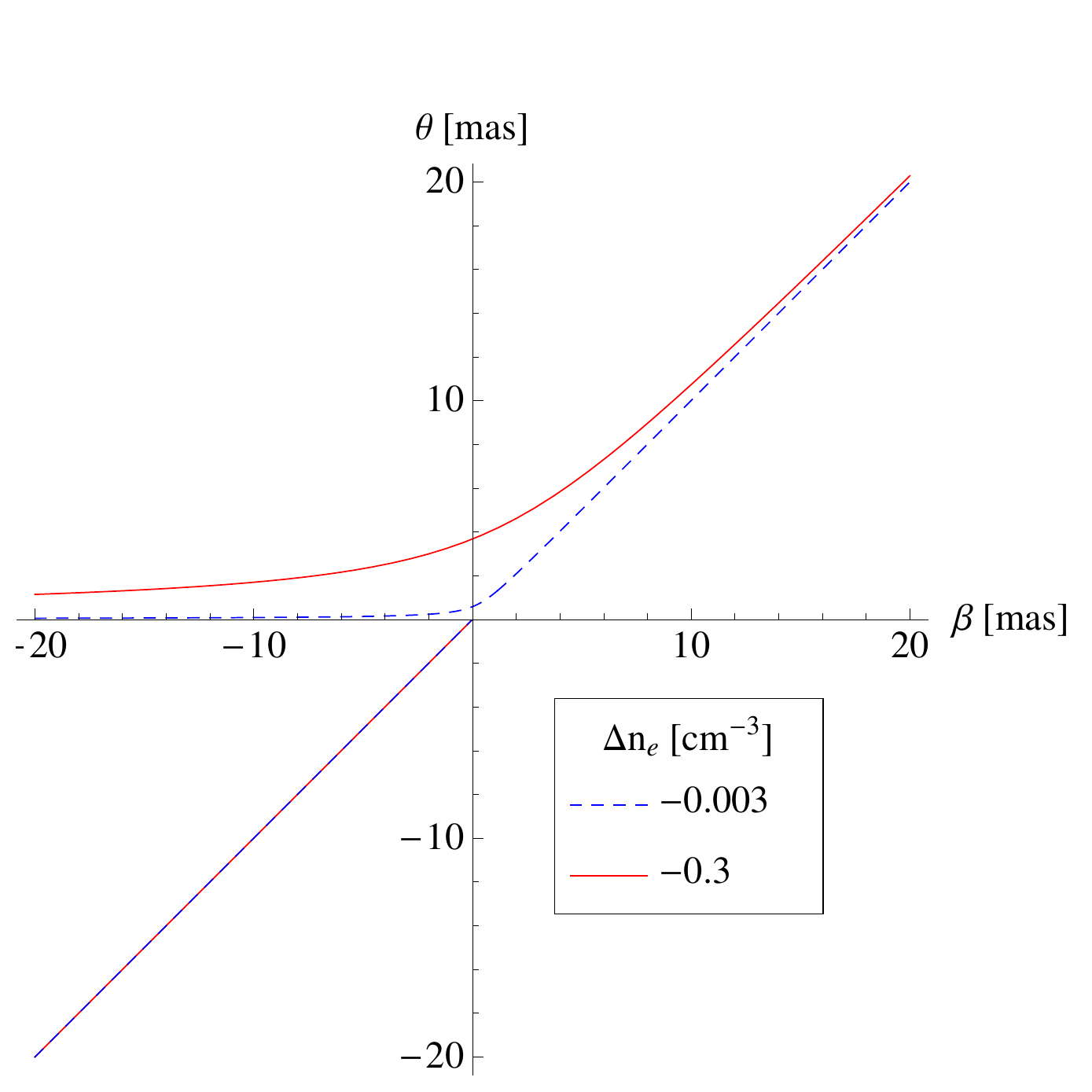}}
 \subfloat[Overdense lens \label{fig:lens_divergent}]{
    \includegraphics[width=0.5\textwidth]{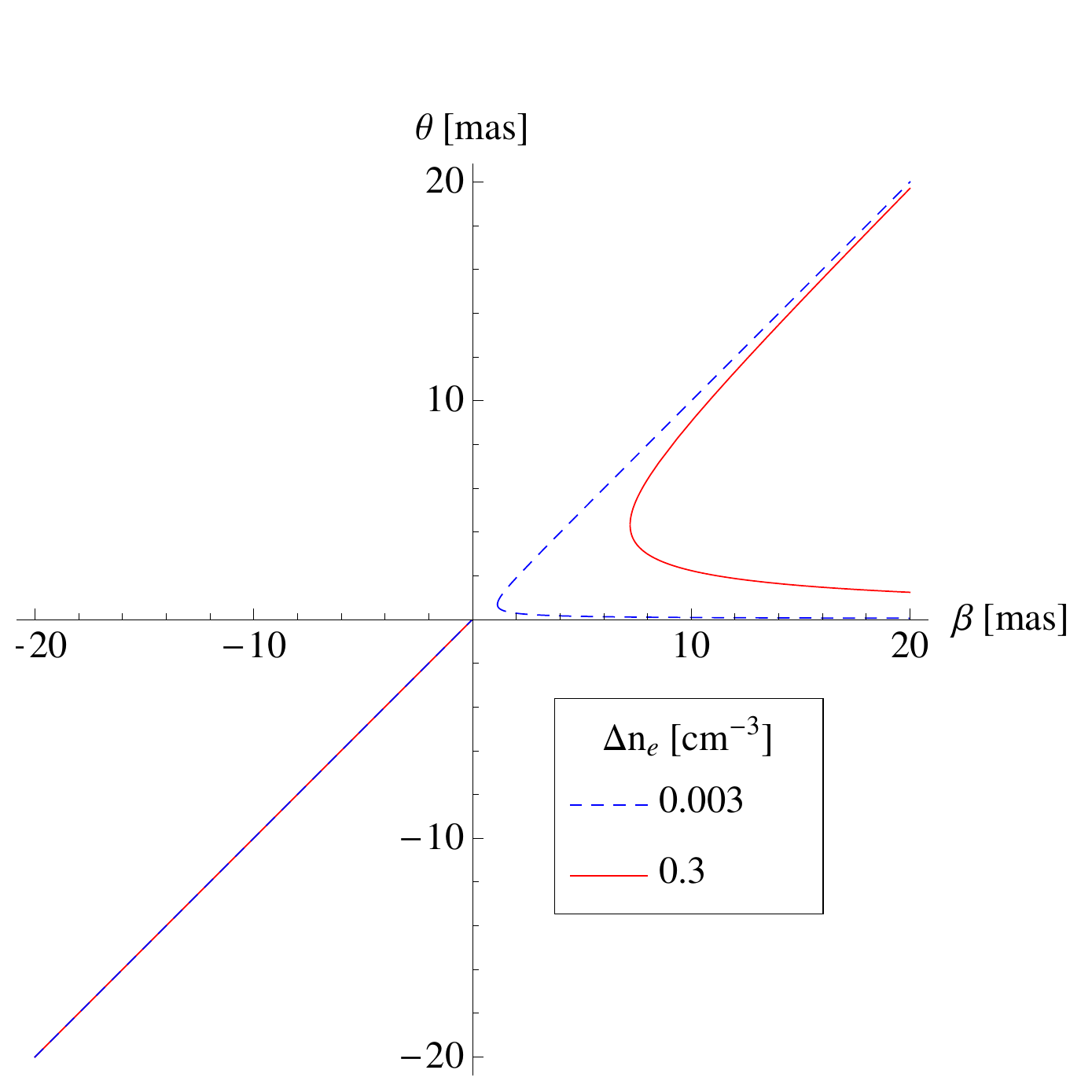}}\\
  \subfloat[Underdense lens\label{fig:magnification_convergent}]{
  \includegraphics[width=0.5\textwidth]{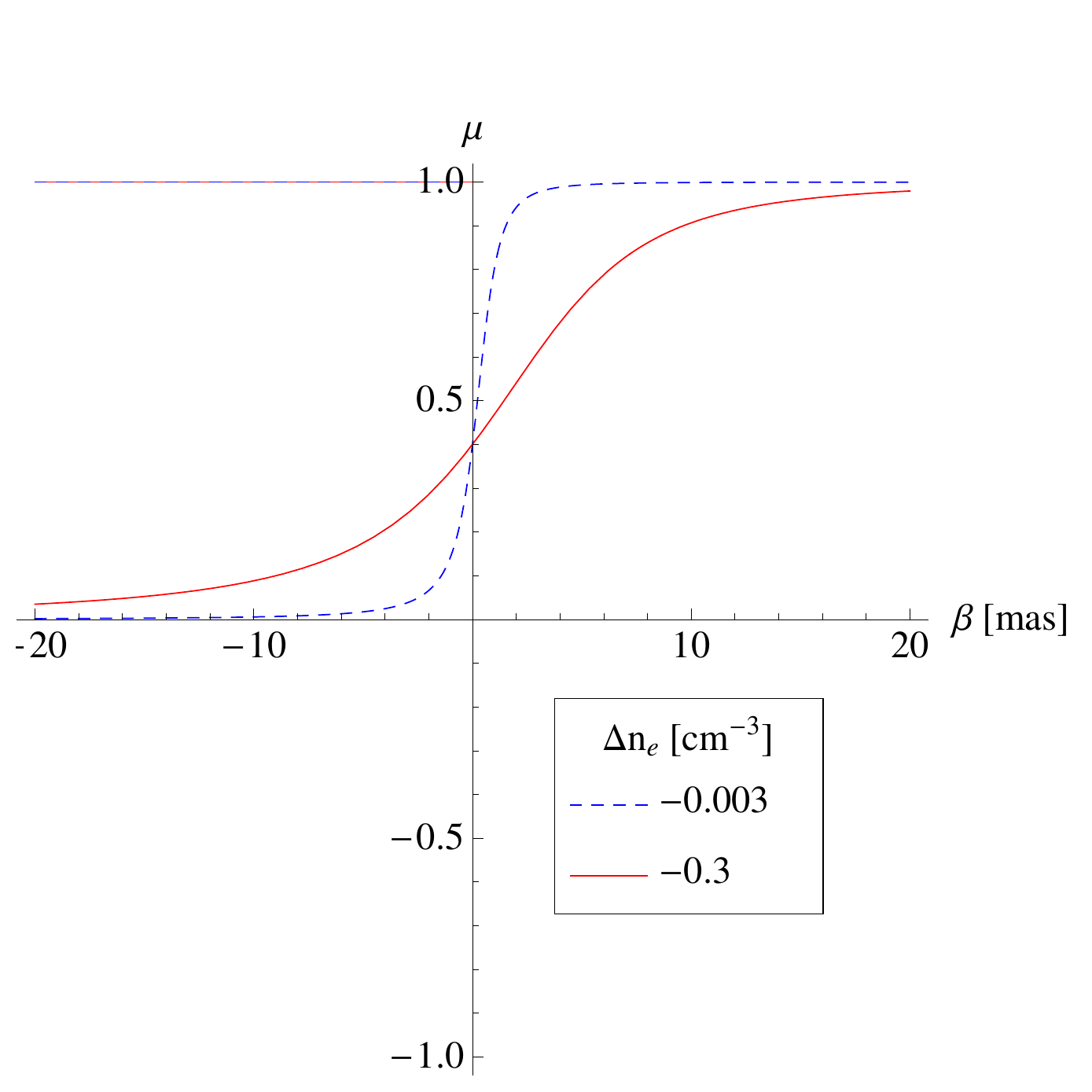}}
 \subfloat[Overdense lens\label{fig:magnification_divergent}]{
    \includegraphics[width=0.5\textwidth]{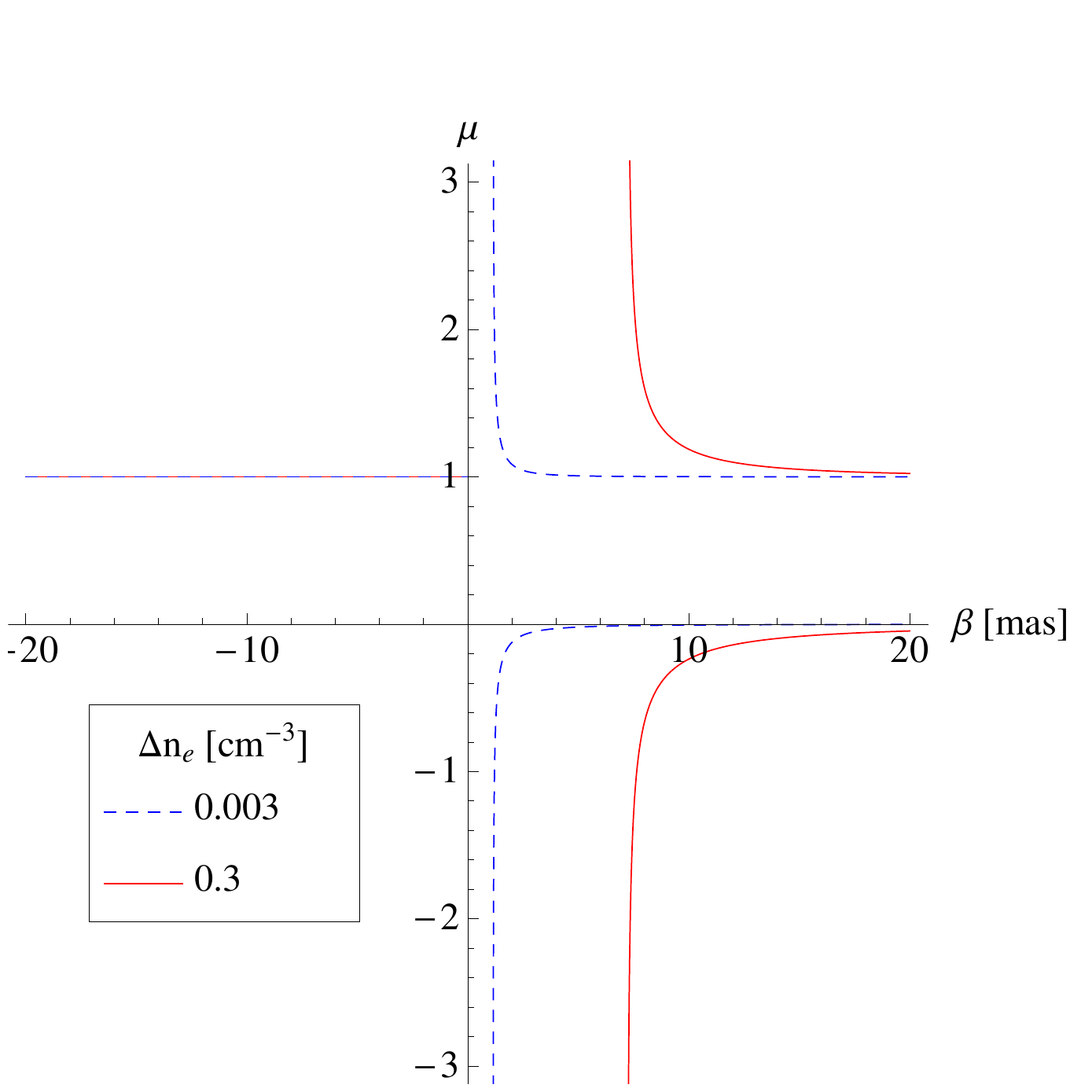}}
  \caption{  
  The observed position of the pulsar, $\theta$, and magnification, $\mu$, as a
    function of the true position, $\beta$.   The legend indicates the difference between
    the free  electron density inside and outside of the lens, $\Delta n_e$ when
    the parameters in Section \ref{sec:parameters} are assumed.  Recall that the lens we
    are considering is at $\theta > \theta_T/2$, or $\theta > 0.04$ mas.
    In the
    underdense lens case, a lensed image is observed along with the true, 
    unlensed image 
    of the pulsar
    before the pulsar passes behind the crest, as the lens
    is bending light from the pulsar into our line-of-sight.
    The 
    angular
    separation between these two images reaches a minimum, non-zero value
    just as the pulsar passes behind the crest ($\beta = 0$), and the
    magnification of the lensed image at this point is $\frac{2}{5}$,
    independent of the lens parameters.  After the 
    pulsar passes
    behind the crest, the line-of-sight image to the pulsar is obscured by the lens, and
    a single, lensed image is observed.  
        In the case
    of an overdense lens, only the line-of-sight image is seen before the pulsar is
    passing behind the crest as the lens bends light out of our line-of-sight.  
    Just after the pulsar has passed behind the lens, no images are seen, as the line-of-sight image
    is obscured by the lens, and the lens is still bending light out of our 
    line-of-sight.
    There is a minimum $\beta$ for which two
    images are produced, and at which the two images have zero angular 
    separation and are both strongly magnified.  The angular separation between the images grows as the pulsar moves 
    to larger $\beta$. The unlensed image is at
    $\theta = \beta$ and $\mu = 1$ for all lenses when $\beta <
    -\theta_T/2$ .
}\label{fig:magnification}
\end{figure*}

Using equations \eqref{eqn:lens} and \eqref{eqn:magnification}, we can
calculate the angular positions and magnifications of images for
lenses of varying strengths, shown in Fig. \ref{fig:magnification}.
When 
examining these figures, recall that the part of the lens we are considering
is at $\theta \ge \theta_T/2$, with maximum strength at
$\theta=\theta_T/2$. 
From Fig.  \ref{fig:magnification}, we can confirm 
that multiple images are produced by the underdense lens when the pulsar is
not behind the crest, and multiple images are produced by the overdense
lens when
the pulsar is behind the crest.  We also note that
magnifications larger than 1 occur only
in the overdense case.  If the lens is overdense, the fainter image is inverted,
as indicated by the negative magnification, and there is a range  
in $\beta$ for which we see no images at all as the 
line-of-sight to the pulsar is obscured by the lens and the lens is
bending light out of our sight.   

\begin{figure*}
  \centering
  \includegraphics[width=0.5\textwidth]{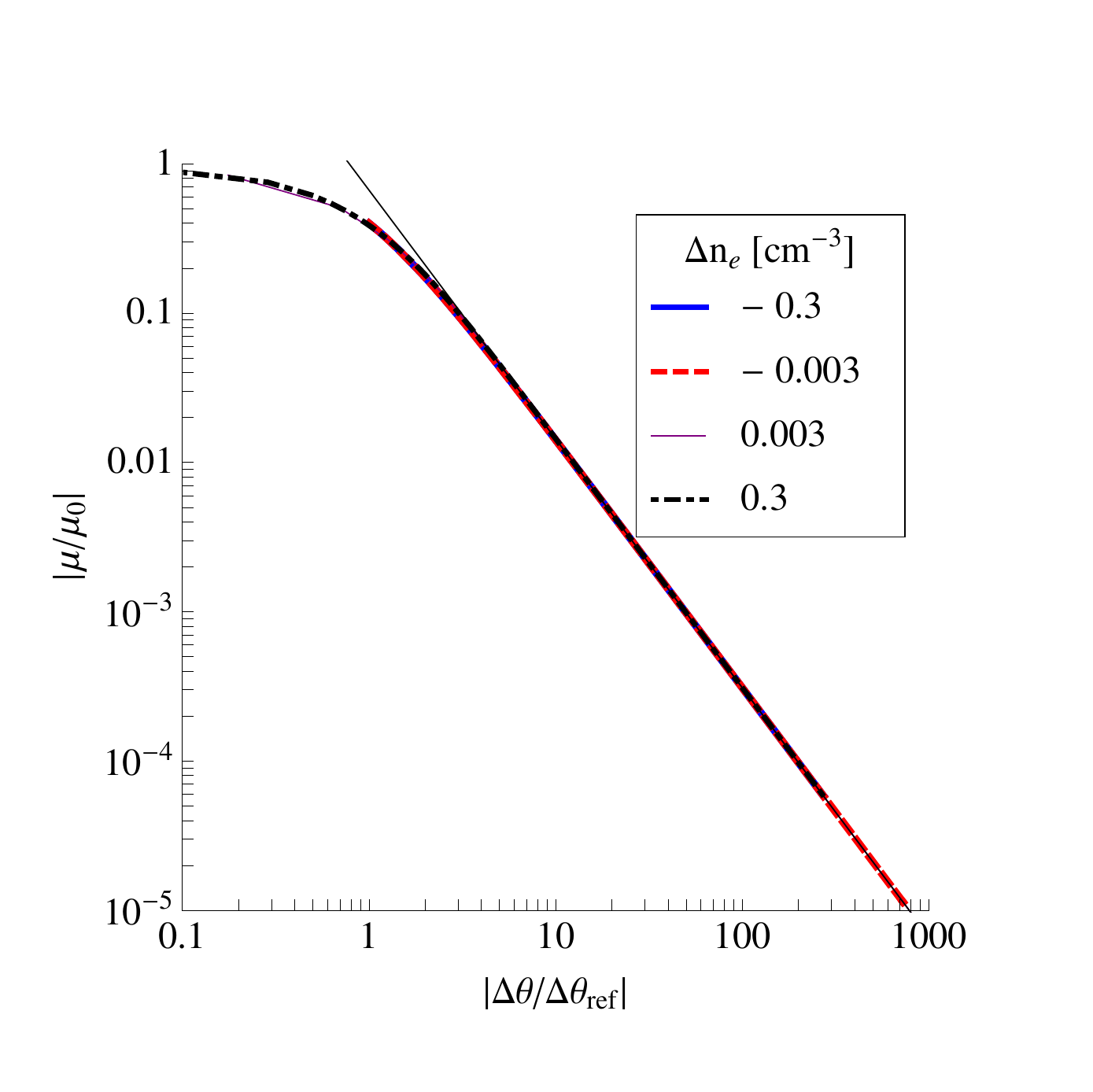}
  \caption{ The ratio of the magnification of the fainter image, $\mu$, to the
    magnification of the brighter image, $\mu_0$, as a function of the angular
    separation, $\Delta \theta$, between two images formed at a single wave crest, scaled by the
    reference value, $\Delta \theta_{\mathrm{ref}}$ (equation \eqref{eqn:dthetaref}).  The legend indicates the difference between the
    electron density inside and outside the lens, $\Delta n_e$ when
    the parameters in Section \ref{sec:parameters} are assumed.  
    The reference separations are 0.58 and 3.68 mas for $\Delta n_e$ of  $\pm 0.003$ and
     $\pm 0.3$ cm$^{-3}$ respectively, and the solid black line shows the analytical relation at large separations, equation \eqref{eqn:magnification5}.  After
    these shifts in angular separation, the curves for various values
    of the lens parameter lie directly on top of one
    another, although the curves for an overdense sheet extend to smaller angular separations due to the minimum angular separation in the underdense case.  This allows one 
    to measure the 
    parameter describing the lens, $\Delta n_e \theta_T \sqrt{r}$, up to the sign of $\Delta n_e$, from
    measurements of $|\mu/\mu_0|$ 
    and $\Delta \theta$ by determining the reference separation
    required to align the point with the curves.  Further observations, such as the change in
    angular separation with wavelength, can then be used to determine
    whether the lens is overdense or underdense. 
}\label{fig:magnificationvdtheta}
\end{figure*}

The numerical relation between the angular separation and flux ratio
of two images formed at a single wave crest is shown in 
Fig. \ref{fig:magnificationvdtheta}.  
We see that 
after each curve is scaled by the reference separation, $\Delta
\theta_{\mathrm{ref}}$ (equation \eqref{eqn:dthetaref}), the relation
between magnification ratio and angular separation is independent of the lens parameters.   
By measuring the angular separation and 
flux ratio for a pair of images, we can determine the value of $\Delta
\theta_{\mathrm{ref}}$ required to place the point along the curves in
Fig. \ref{fig:magnificationvdtheta}, and therefore 
$| \Delta n|\, r \sqrt{\theta_T}$, provided that the distance to
the pulsar is known.  

\subsection{Time and frequency evolution of the system}
\label{sec:evolution}

Once the free parameter of the lens, $|\Delta n| \, \theta_T \sqrt{r}$,
is constrained from the magnification ratios and angular separations,
changes in the locations and 
magnifications of the 
images with frequency can be predicted.  If the angular velocity
of the pulsar relative to the screen in the $x$-direction (parallel to
the axis along which the images are formed; see
Fig. \ref{fig:geometry}),
$\frac{(V_{\mathrm{psr}}-V_\mathrm{lens})_\parallel}{d_\mathrm{psr}}$,
is also known, we can  
predict changes over time.  Since we do not know if the
lens is overdense or underdense at this stage, we 
consider both cases.

As an example, we will consider two images with a magnification ratio of 0.01
and an
angular separation of 10 mas at a frequency of 314.5 MHz.  Using these
values, we calculate that the  
lens has, 
for the parameters given in Section \ref{sec:parameters}, $|\Delta
n_e| = 0.007$ cm$^{-3}$.  From the lens equation and the angular
separation,  
we now determine the true position of the
pulsar relative to the crest, $\beta$, for the underdense 
and overdense cases, -9.9 and 10.2 mas respectively.  
Let's assume that the angular separation is decreasing as the pulsar moves
towards the crest.  
In the case of the overdense lens, after 4 weeks, the
magnification ratio of the 2 images has increased to 0.028, and the
angular separation has decreased by 4.6 mas. 
The magnification ratio of the images will continue to increase  and
the images will continue to get brighter until
the two images have zero angular separation after 7.6 weeks, after 
which the lensed image will disappear.  
In the case of the
underdense lens, after 4 weeks, the angular separation between the two
images has decreased by 4.4 mas, and the magnification of the lensed image is
0.026.  

Similarly, we can consider changes with frequency.
Using the same example, we examine how the angular separation between
the two images varies over a 32-MHz frequency band from 310.5 MHz to
342.5 MHz and find that, for the underdense lens, the angular
separation will decrease by 0.019 mas moving from the lowest to
highest frequencies in the band, while the magnification ratio
decreases by 0.0012.  For the overdense lens, the angular separation
increases by 0.022 mas from the lowest to highest frequencies in the
band, while $|\mu/\mu_0|$ decreases by 0.0013. 
We fit a power law
to the relation between angular separation and wavelength over this
band for both the underdense and overdense lenses.  The best fit power law
exponents are shown
in Table \ref{tbl:wavelength}.
In the underdense case,
the power law exponent is positive, meaning that at higher frequencies
the images are closer together, while in the
overdense case it is negative, so that at higher frequencies the
images are further apart.  This is also apparent from Figs. \ref{fig:lens_convergent} and \ref{fig:lens_divergent}:  Since $\Delta n \propto - \Delta n_e\, \lambda^2$, observing at longer wavelengths has the same effect as increasing the strength of the lens, and one can see from Figs. \ref{fig:lens_convergent} and \ref{fig:lens_divergent} that when the lens is stronger (or at lower frequencies) the angular separation is larger if the corrugated sheet is underdense, but smaller if the corrugated sheet is overdense.
\citet{brisken_100_2010} statistically combine arclets in the
secondary spectrum of PSR B0834+06 and measure positive power law 
exponents, $0.062 \pm 0.006$ and $0.019 \pm 0.004$, over this 32-MHz
band for two of three identified arclet groups 
(for the third group, they find that the separation does not vary with
wavelength within the measurement uncertainties), suggesting that 
the sheet is underdense and indicating that the small
exponents predicted by this model are measurable. 

\begin{table}
  \caption{Power law fits of the form $\Delta \theta = B
    \lambda^{\gamma}$ to the change in separation with
    wavelength for a lens that produces two images with an angular
    separation of 10 mas and a flux ratio of 0.01 at 314.5 MHz.  The
    fit is done over 310.5 MHz to 342.5
    MHz.  While the exponents here are small, exponents of
    the same order of magnitude have been measured over the same
    bandwidth with 10 to 20\% uncertainties by
    \citet{brisken_100_2010}.
}
  \label{tbl:wavelength}
\begin{tabular}{cc}
  \hline
  $\Delta n_e$ [cm$^{-3}$]  & $\gamma$ \\ \hline
   -0.0067  & 0.019 \\ 
  0.0067  & -0.023 \\ \hline 
\end{tabular}
\end{table}

\section{Measurements of Pulsar Scattering}
\label{sec:observations}

The main phenomenon this model hopes to reproduce is the existence of
parabolic arcs in the secondary spectra of pulsars, which can arise
from highly anisotropic scattering at a thin sheet along 
our line-of-sight to the pulsar.  Using global VLBI to obtain the
secondary cross-spectrum between 
two stations, the angular locations of the images relative to
the line-of-sight image of the pulsar can be measured \citep{brisken_100_2010}.  Under the
thin sheet approximation, the 
distance and velocity of the sheet can be determined from the
relationship between the measured angular separations of two images 
and the Doppler frequency and delay
of the feature in the secondary spectrum resulting from the interference 
between these images.
If inverted
arclets, evidence of distinguishable images on the screen, are present
in the secondary spectrum, the location of these in the secondary
spectrum combined with the distance and velocity of the sheet can be
then used to obtain the positions of the images to 
$\approx 1\%$ accuracy \citep{brisken_100_2010}, a technique known as
`back-mapping'.  Using additional 
techniques, such as holography, the precision of these measurements can
be improved 
even further \citep{pen_50_2014}. The relative fluxes of inverted
arclets in the secondary spectrum encodes the relative fluxes of the
images of the pulsar on the scattering screen, providing us with the
rest of the information required to constrain the model.  To test temporal
evolution of the scattering against this model, we also need to know
the whether the angular separation between an image and the
line-of-sight image to the pulsar is
increasing or decreasing.  If the Doppler frequency of the arclet
corresponding to an image is
negative, then the pulsar is
moving towards the image and the angular separation
is decreasing.  If the Doppler frequency is positive, then the pulsar
is moving away from the image, and the angular
separation is increasing.  

This model may also be applied to pulse echoes, such as those
observed in the Crab pulsar
\citep[eg.][]{backer_plasma_2000,lyne_pulsar_2001}, PSR B2217+47 \citep{michilli_echoes_2018}
and at least one other pulsar (Oslowski \textit{et al.}, in prep.).  In these cases, 
radio interferometry can be used to measure the angular position of the echo
relative to the main pulse,
while the pulse profile itself can be used to determine the
magnification of the echo.  Combining the angular information with the
delay of the echo relative to the main pulse allows one to determine
the geometry and velocities (if one has multi-epoch observations of
the system) of the scattering system.  
By correlating the unlensed pulse with the echo, the phase imparted by 
the lens can be retrieved.  If the image is inverted, the
waveform of the image will be distorted \citep{dai_waveforms_2017}, 
allowing a direct test for the inversion of the image in this case.
Some pulsars have giant pulses,
short, intense bursts of radiation,
that exhibit scattering tails from interstellar scattering.  Giant
pulses allow the response of the lens to be determined from the
scattering of a single pulse
\citep{main_descattering_2017}, making these systems even more
promising for detecting image inversion.

\section{Extensions}
\label{sec:extensions}

\subsection{ The region of the lens $\mathbf{-\theta_T/2  < \theta < \theta_T/2}$}
\label{sec:innerlens}

So far, we have considered only the portion of the crest where $\theta
> \theta_T/2$.  However, we expect additional images from the region
$-\theta_T/2 < \theta < \theta_T/2$
The maximum extent of the
lens along the 
line-of-sight occurs at
$\theta=\theta_T/2$, so that the gradient $\nabla_x Z$ has opposite sign for 
$|\theta|<\theta_T/2$ and $\theta>\theta_T/2$, and therefore the
bending angle, which is proportional to  
this gradient, switches direction at $\theta=\theta_T/2$.  

We can treat the region $-\theta_T/2<\theta<\theta_T/2$ as a separate lens, and we can parameterize the lens as being 
bounded by the line 
\begin{equation}
z = \sqrt{(R+T)(x+T/2)} \;.
\end{equation}
The thickness of the lens is then
\begin{equation}
Z = 2 \sqrt{(R+T)(x+T/2)}\;,
\end{equation}
and, once again assuming the index of refraction is constant within the lens, the bending angle induced by the lens is
\begin{align}
\hat{\alpha}(x) &= - \Delta n \nabla_x Z \\
                               &= - \Delta n \sqrt{\frac{R+T}{x+T/2}} \\
\hat{\alpha}(\theta) &= - \Delta n \sqrt{\frac{r+\theta_T}{\theta + \theta_T/2}} \;,
\end{align}
so that the lens equation for this system is:
\begin{equation}
\theta = \beta - s \Delta n \sqrt{\frac{r+\theta_T }{\theta + \theta_T/2}} \;.
\label{eqn:innerlens}
\end{equation}
From equation \eqref{eqn:innerlens}, we see that the region $-\theta_T/2 < \theta < \theta_T/2$ corresponds to 
$\beta \ge \frac{\theta_T}{2} + s |\Delta n| \sqrt{\frac{r+\theta_T}{\theta_T}} $ if the lens is underdense, and to $\beta \le \frac{\theta_T}{2}- s | \Delta n | \sqrt{\frac{r+\theta_T}{\theta_T}} $ if the lens is overdense.

Once again, we can determine the magnification of the image produced from the derivative of the 
lens equation:
\begin{align}
\mu &= \bigg( \frac{\mathrm{d}\beta}{\mathrm{d}\theta}\bigg)^{-1} \\
&= \bigg( 1-\frac{s \Delta n \sqrt{(r+\theta_T)/4}}{(\theta+\theta_T/2)^{3/2}} \bigg)^{-1} \;. 
\end{align}
We see that if $\theta_T \gtrsim (s |\Delta
n| \sqrt{(r+\theta_T)/4})^{2/3}$, a very highly magnified image can occur in the region $ -\theta_T/2 <\theta < \theta_T/2$ and would need to be considered.
However, if $\theta_T \ll (s |\Delta
n| \sqrt{(r+\theta_T)/4})^{2/3}$, which holds for the parameters in Section \ref{sec:parameters}, the magnification is instead maximized when $\theta = \theta_T/2$, and reaches a value 
\begin{equation}
|\mu|_\mathrm{max}  \approx \frac{\theta_T^{3/2}}{s |\Delta n| \sqrt{(r+\theta_T)/4}} \;.
\label{eqn:inner_mumax}
\end{equation}
For the parameters in Section \ref{sec:parameters} and $\Delta n_e = 0.3$ cm$^{-3}$, $|\mu|_\mathrm{max} = 9\mathrm{x}10^{-5}$, 
much fainter than the images produced
at $\theta > \theta_T/2$.  We see from equation \eqref{eqn:inner_mumax} that 
if $\theta_T \ll (s |\Delta
n| \sqrt{(r+\theta_T)/4})^{2/3}$, the images produced by
the region $-\theta_T/2 < \theta < \theta_T/2$ are very faint compared to the unlensed image.

The constraints on the angular positions of the pulsar at which this faint
image appears result in a minimum angular separation between the faint image 
and the bright image, or line-of-sight image, of the pulsar,
\begin{align}
|\Delta \theta|_\mathrm{min} &\simeq \theta_T/2 - \beta(\theta_T/2) \\
                             &= s | \Delta n| \sqrt{\frac{r+\theta_T}{\theta_T/2}} \;.
\end{align}
This is exact when the sheet is overdense, and the faint image from 
$-\theta_T/2 < \theta < \theta_T/2$ appears simultaneously with the line-of-sight
image to the pulsar.  If the sheet is underdense, the faint image appears 
simultaneously with the lensed image at positive $\beta$, but as this faint image 
from $-\theta_T/2 < \theta < \theta_T/2$ appears only at large $\beta$, where the 
bright lensed image from $\theta > \theta_T/2$ appears at $\theta \approx \beta$, 
this approximation is very good.  For the parameters in Section \ref{sec:parameters} and $\Delta n_e = 0.3$ cm$^{-3}$, 
$|\Delta \theta|_\mathrm{max} \approx 2000$ mas.  In contrast, in the secondary spectrum 
of PSR B0834+06, scattered flux appears up to an angular separation of 28 mas.  Since 
this is less than the minimum angular separation between the line-of-sight image of the
pulsar and the faint image from $-\theta_T/2 < \theta < \theta_T/2$, we see once again that 
we can ignore the images from this region of the lens.  At larger thicknesses, the lensed images 
become brighter and move to lower angular separations, where they can complicate the 
analysis and should be considered.  

Due to the change in lensing direction at $\theta = \theta_T/2$, we may 
be able to place independent constraints on the
thickness of the sheet.
Consider the case where a faint
image from a region $\theta >\theta_T/2$ is moving towards
smaller $\theta$ as the pulsar moves away from the crest. 
Once the image reaches $\theta = \theta_T/2$, if the pulsar continues
to move further from the crest, the lens 
will no longer be able to bend the pulsar light into our
line-of-sight, and the faint image and corresponding echo or arclet in the
secondary spectrum will disappear.
This sets a maximum angular separation between the two images when
$\theta_{\mathrm{max sep.}} = \theta_T/2$.
If this
maximum angular separation could be observed, then from the angular
separation and magnifications of the images
we could determine the parameter of the lens, the true position,
$\beta$, and the lensed position, $\theta_{\mathrm{min}}=\theta_T/2$
from $\Delta \theta_\mathrm{ref}$ (\textit{e.g.} Section \ref{sec:results}).
If we know the distance to the lens, we can also calculate the 
physical thickness of the lens, $T$.  For the parameters in
Section \ref{sec:numerical} and $\Delta n_e = 0.3$ cm$^{-3}$, the maximum angular separation is 3000
mas, at which the flux ratio
between the two images is $-8\mathrm{x}10^{-6}$.  Typical secondary spectra analyses are not sensitive to angular separations larger than approximately 30 mas, so the maximum angular
separation is not expected to be observable as a sudden disappearance
of the lensed image. 

\subsection{A smooth lens profile}
\label{sec:smoothed}
In previous sections, we have considered the effects of a lens profile with a discontinuous derivative.  
This is due to our 
assumption that the density within the sheet is constant, so that the density 
does not vary smoothly across the sheet.
We can relax this assumption with a more general treatment, with which we can consider any density profile through the sheet, 
$\Delta n_{e,\mathrm{sheet}}(d)$, where $d$ is the depth through the unbent sheet, normal to the plane of the sheet.

We write the bending angle, $\alpha$ as:
\begin{equation}
\alpha = - s \frac{\lambda^2}{2\pi}r_e \nabla_x \Delta N_e(x) \;,
\end{equation}
where $\Delta N_e(x) = \int_{-\infty}^\infty dz (n_e(x,z) - n_{e,0})$ and $n_{e,0}$ is the free electron density in the ambient ISM.
We can approximate $N_e(x)$ as the convolution of the profile of the corrugated sheet and the electron density profile through the sheet, $\Delta n_{e,\mathrm{sheet}}(d)$:
\begin{equation}
\Delta N_e(x) = \int_{-\infty}^\infty \mathrm{d}X \, \Delta n_{e,\mathrm{sheet}}(X) \, 2 \frac{\mathrm{d}l}{\mathrm{d}x}\bigg|_{x-X} \;.
\end{equation}
This is a good approximation when the tangent to the corrugated sheet is near parallel with our line of sight to the pulsar, such as near the crest of a wave on the sheet.
In order to compare between different density profiles, we normalize the electron density profile through the sheet so that
\begin{equation}
\int_{-\infty}^\infty \mathrm{d}d\,\Delta n_{e,\mathrm{sheet}}(d) = T \Delta n_e \;,
\end{equation}
where $T=0.03$ AU and $\Delta n_e = 0.3$ cm$^{-3}$.  

We consider a Gaussian form of the electron density profile,
\begin{equation}
\Delta n_{e,\mathrm{sheet}}(d) = \frac{\Delta n_e T}{\sqrt{2 \pi} s} \, \exp\bigg( -\frac{d^2}{2 s^2}\bigg) \;,
\end{equation}
where $s = \frac{T}{2 \sqrt{2 \ln(2)}}$ and $\Delta n_{e,\mathrm{sheet}}(d)$ has a FWHM of $T$, as an example of a smoothly-varying density profile and contrast this with a top-hat electron density profile, 
\begin{equation}
 \Delta n_{e,\mathrm{sheet}}(d) = \Delta n_e\, \Theta(T/2-d)\,\Theta(T/2+d)\;,
\end{equation}
where $\Theta$ is the Heaviside step function.  This top-hat electron density profile, allows us to extend the model discussed in previous sections of this paper, where we investigated a top-hat electron density profile using analytic approximations, to the entire lens, including those regions where our previous approximations did not hold.

\begin{figure*}
  \centering
  \includegraphics[width=0.5\textwidth]{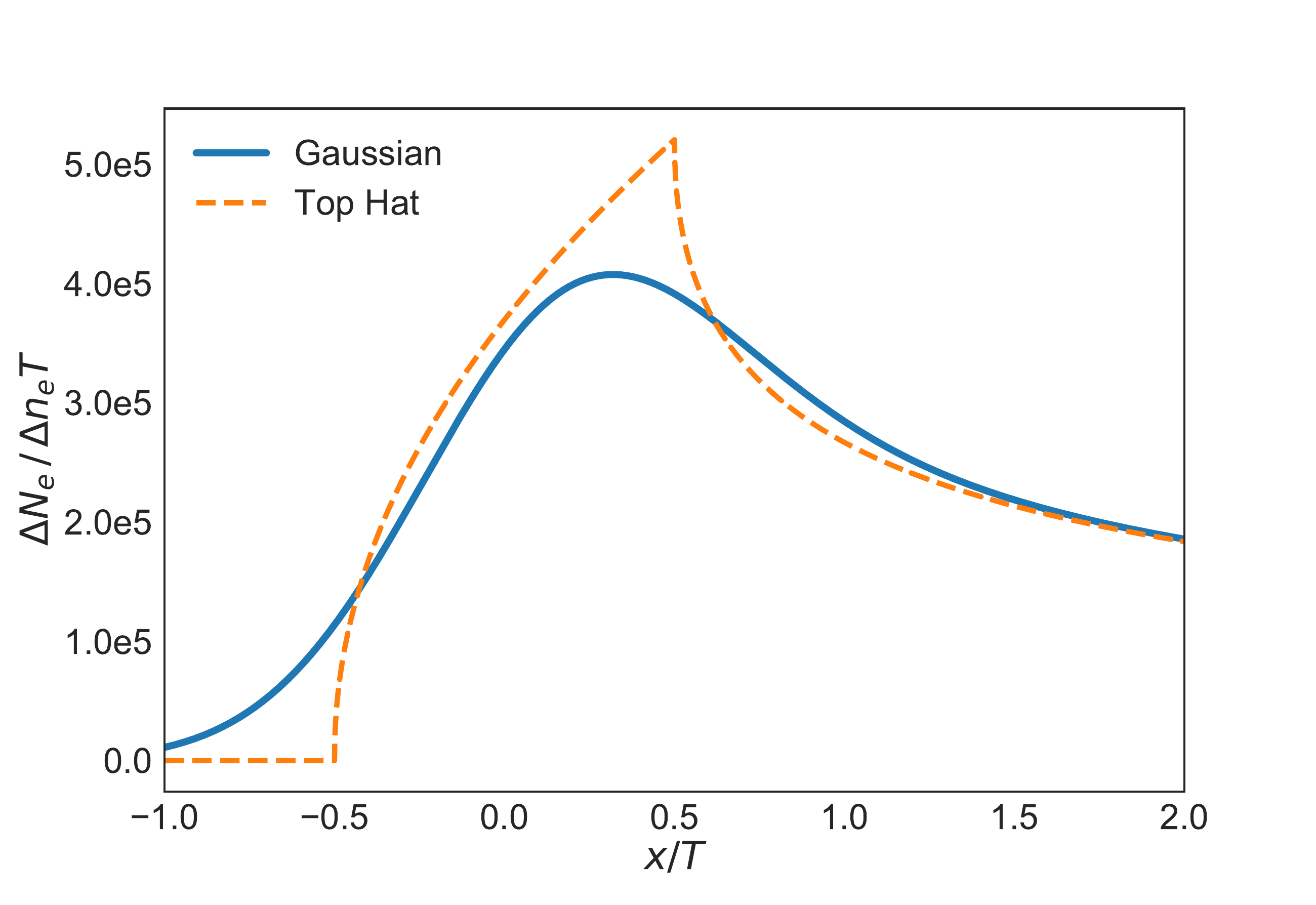}
  \caption{The lens profiles as a function of $x$ for a top hat (orange dashed line) and a Gaussian (blue solid line) electron density profile through the sheet.  Note that the lens profile is continuously differentiable for the Gaussian electron density profile, but that the lens profile from a top hat electron density profile, like that considered analytically in the sections above, has a discontinuous derivative at $x=-T/2$ and $x=T/2$.  As a result, the corrugated Gaussian sheet adheres to the odd-image theorem while the corrugated top hat sheet does not.}\label{fig:smoothprofiles}
\end{figure*}

\begin{figure*}
  \centering
 \subfloat[Underdense sheet\label{fig:underdense_smooth}]{
 \includegraphics[width=0.5\textwidth]{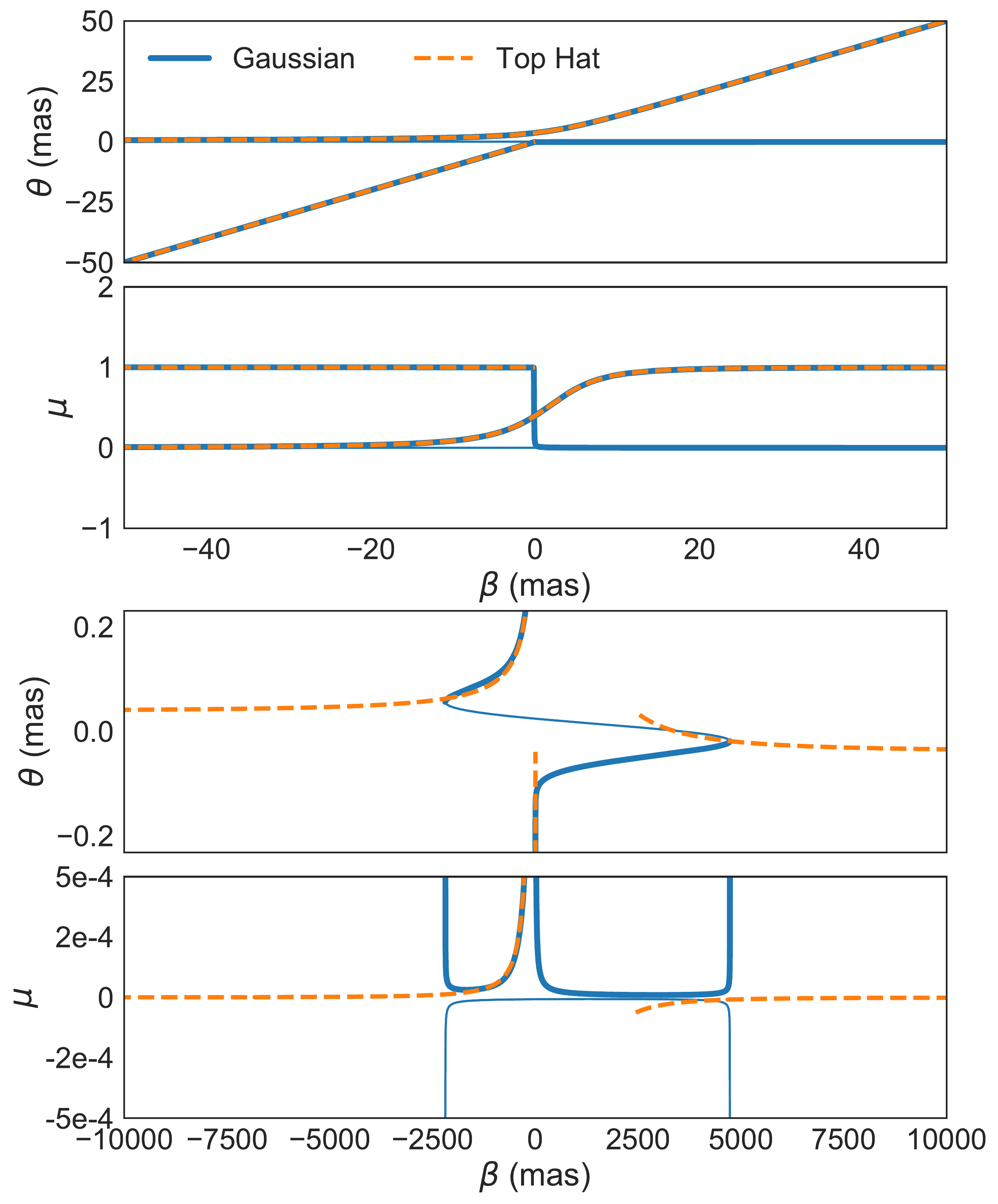}}
 \subfloat[Overdense sheet\label{fig:overdense_smooth}]{\includegraphics[width=0.5\textwidth]{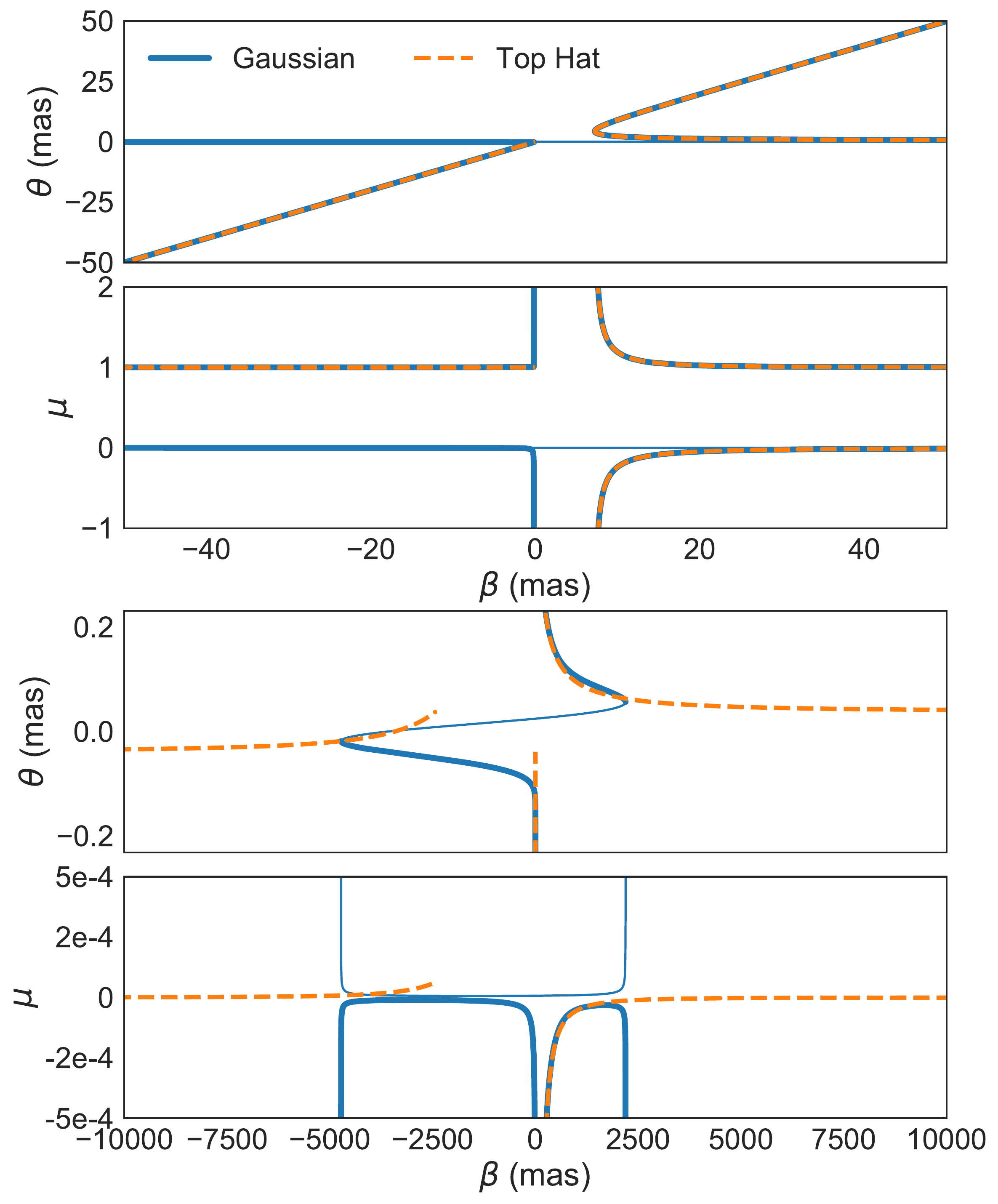}}
  \caption{The solution to the lens equation for the lens profiles in Fig. \ref{fig:smoothprofiles}.   From the top two panels of each subfigure, we see that when $\theta \gg \theta_T$, the lens behaviour does not depend on the electron density profile through the sheet apart from the extra image formed near $\theta = 0$ when the sheet has a Gaussian profile.   In the bottom two panels, we have adjusted the axis to focus on region $|\theta|<\theta_T$ and highlight the differences that arise between the two electron density profiles.  We have drawn the extra image produced when a smooth Gaussian density profile through the sheet is considered with a thin line to distinguish it from the other images.
  }\label{fig:smooth_lens_behaviour}
\end{figure*}

The lens profiles are shown in Fig. \ref{fig:smoothprofiles}, and the solutions to the lens equation, $\theta(\beta)$ and $\mu(\beta)$, are shown in Fig. \ref{fig:smooth_lens_behaviour}, where we see that the behaviour at $|\theta| \gg \theta_T$  (0.08 mas for these examples) does not depend on the specifics of the density profile through the lens.  However, when $|\theta| \lesssim \theta_T$, the bending angles in the case of the top-hat sheet profile diverge.  This is in contrast to the Gaussian sheet, which adheres to the odd-image theorem and produces additional images compared to the top-hat sheet.  These images are normally very faint but are highly magnified when $\frac{\mathrm{d}\beta}{\mathrm{d}\theta}=0$.   These highly-magnified events occur for large bending angles, when the pulsar is far from the crest of the wave, and for only a small range in $\beta$, and thus in time.  Typical secondary spectrum analyses are sensitive to $|\Delta \theta| \lesssim 30$ mas, and therefore do not need to account for these highly magnified images.  In addition, due to the chromatic nature of the lensing, these events occur at a specific $\beta$ for only a small range in frequency, and thus are expected to be much less apparent when observing with a wide bandwidth. 

\begin{figure*}
  \centering
\subfloat[Underdense sheet \label{fig:underdense_nimages}]{ \includegraphics[width=0.4\textwidth]{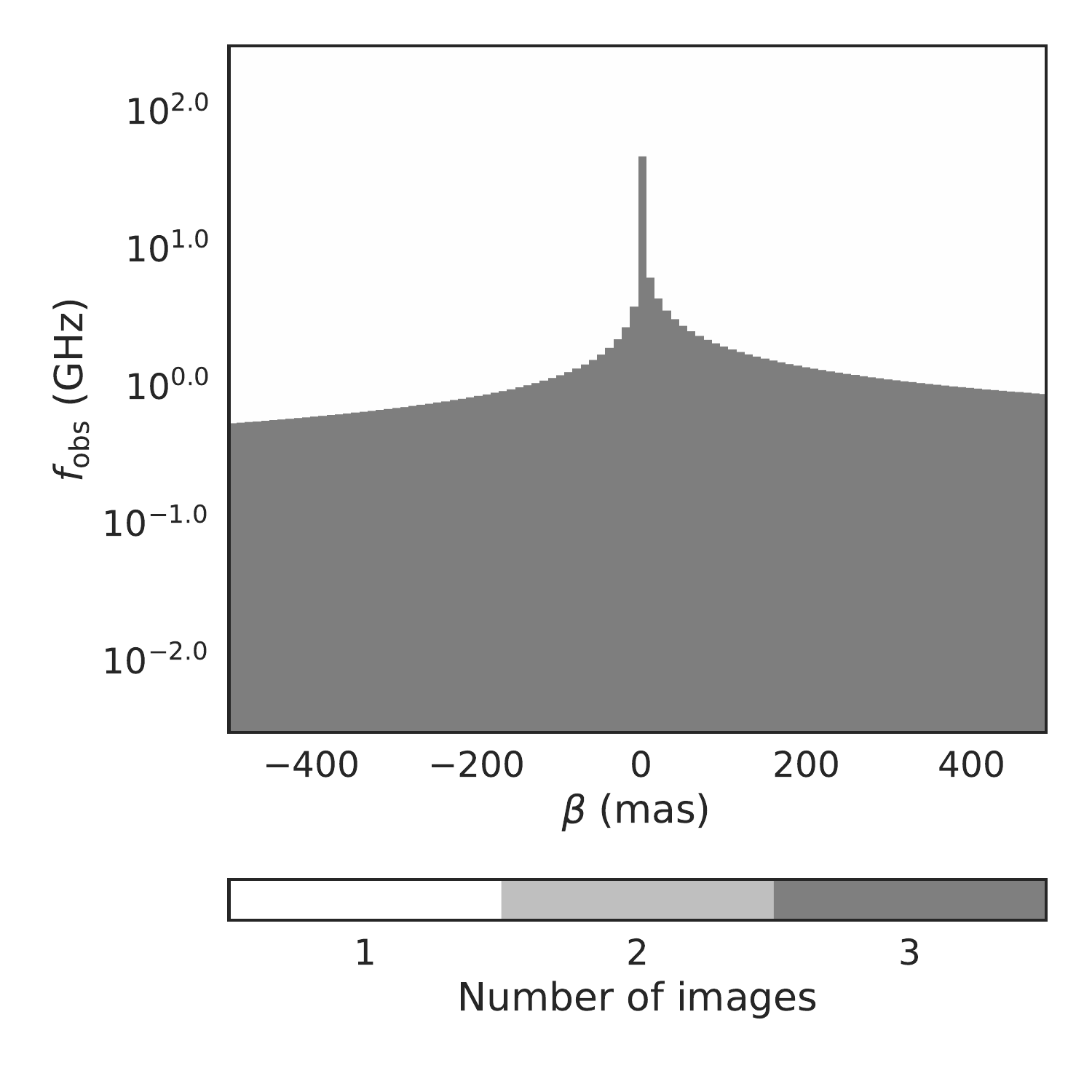}}
\subfloat[Overdense sheet \label{fig:overdense_nimages}]{
\includegraphics[width=0.4\textwidth]{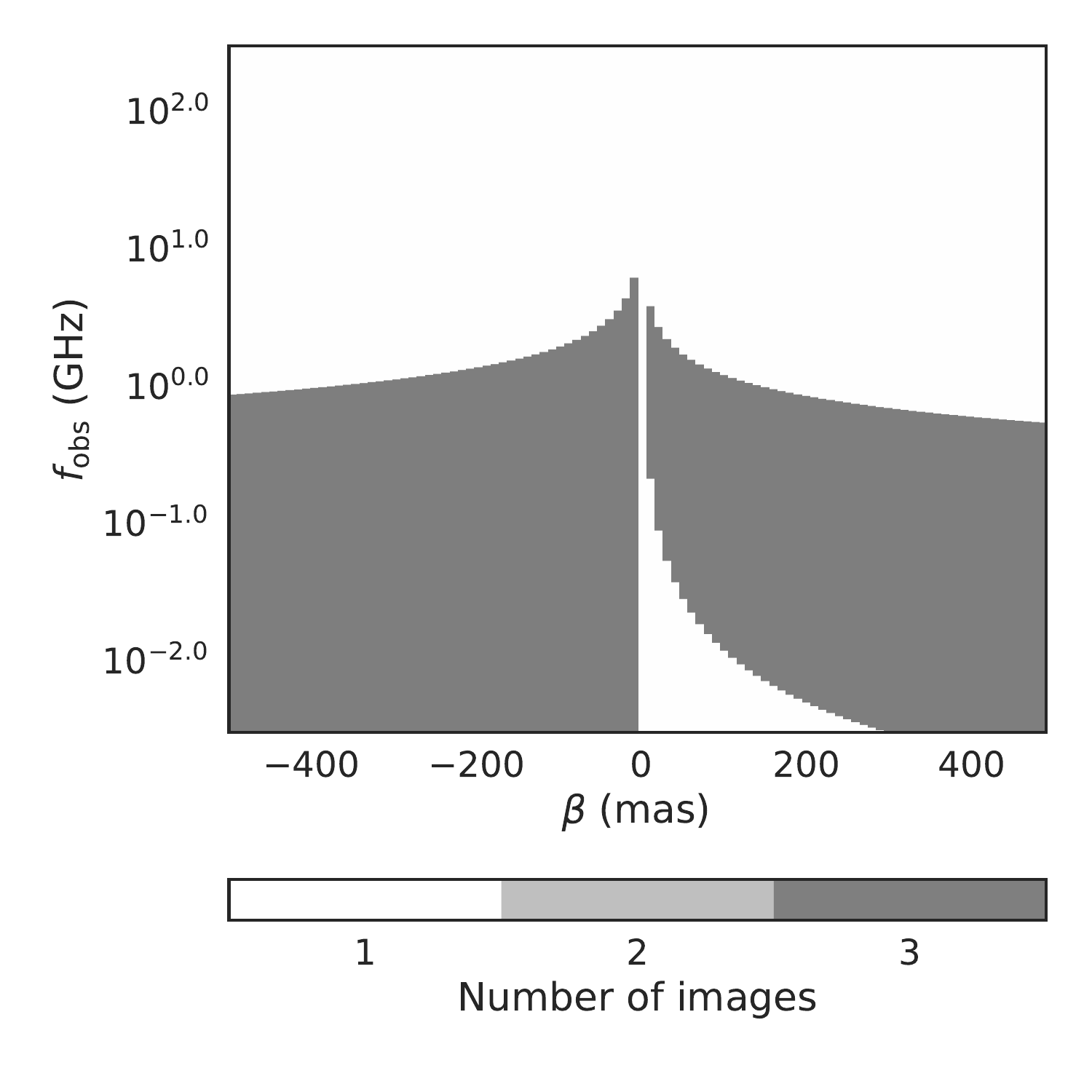}}
  \caption{The dependence of the number of images produced on observing wavelength, $\lambda_\mathrm{obs}$, and the location of the pulsar relative to the crest of the wave, $\beta$, for a sheet with a Gaussian electron density profile.  Both an underdense (Fig. \ref{fig:underdense_nimages}) and overdense (Fig. \ref{fig:overdense_nimages}) sheet are considered.  We see that this model adheres to the odd-image theorem, as there are only regions of 1 or 3 images. }\label{fig:smooth_nimages}
\end{figure*}

We can also look at how the number of images produced by the lensing system varies with both the observing frequency and the offset between the pulsar and the crest of the wave, shown in Fig. \ref{fig:smooth_nimages}.  When observing these images, keep in mind that during the transition between regions of one and three images events of high magnification occur.  As expected from our discussion in previous sections, we see that additional high-magnification events occur in the overdense case compared to the underdense case. 

\subsection{Many images from a single sheet}
\label{sec:multiple}

In practice, we often observe many lensed images of a pulsar.  In
the picture presented in this work, each of these images is from a different wave crest on
the sheet.  Each wave crest may have a different value of $\Delta n \, \theta_T
\sqrt{r}$, and so the analysis above can be done for every wave crest, 
or equivalently every inverted arclet in the secondary
spectrum or every echo of the pulsar.  

While in some pulsars the individual images are distinguishable in the
secondary spectrum as individual
inverted arclets, in many cases these
inverted arclets are not resolved, adding
ambiguity to modeling the lensing in terms of individual wave crests.
Holographic techniques
\citep[\textit{e.g.}][]{pen_50_2014,walker_interstellar_2008}
can retrieve the electric field due to the combination of images,
which may assist in identifying individual images, and we can
additionally consider 
statistical phenomena that  
arise when we have lensing from multiple crests, such as the
the scattering tail of a single pulse, the spread of pulse power over time due to the
delays imparted by the various crests.  This requires numerical
simulations that take into account the distribution of waves along the
sheet, and is deferred to future work. 

\subsection{Symmetric lensing events}
\label{sec:ESEs}

\begin{figure*}
 \centering
  \includegraphics[width=0.2\textwidth]{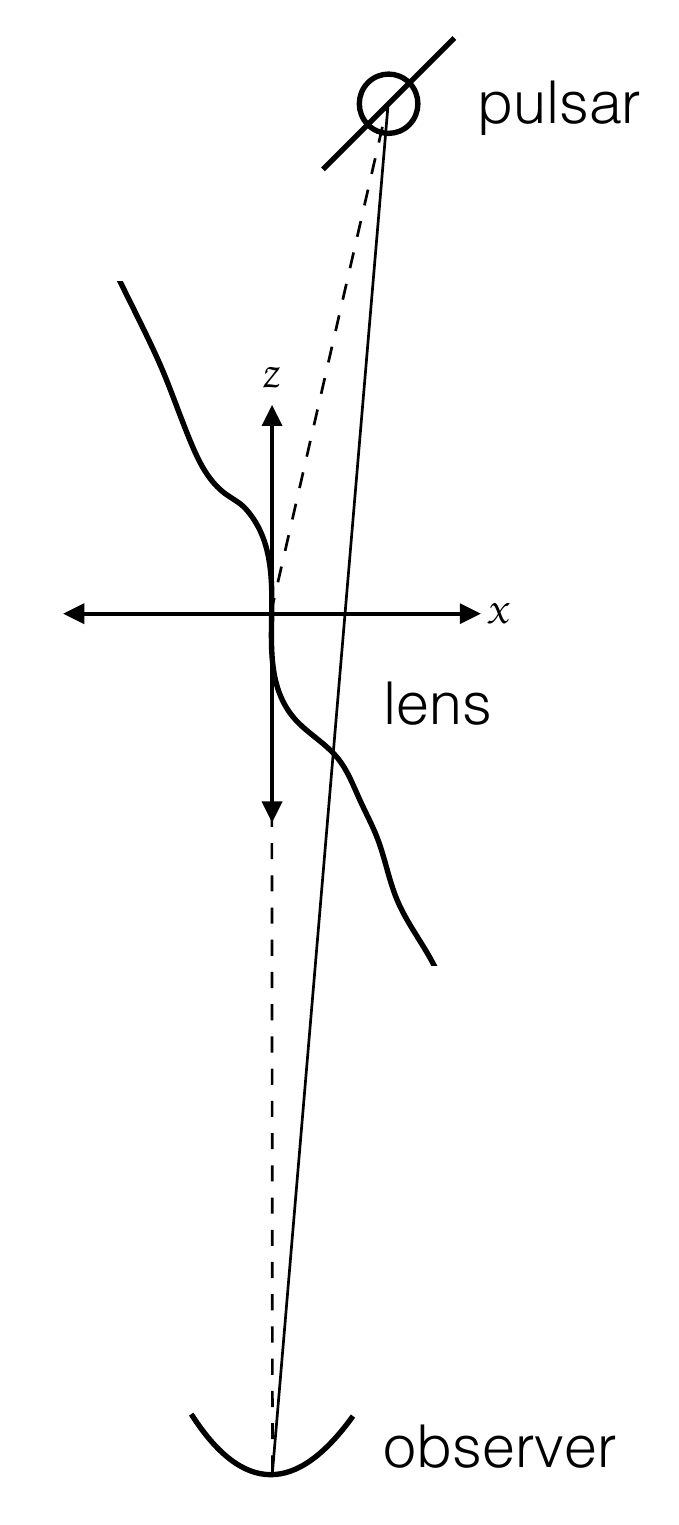}
  \caption{A geometry of the corrugated sheet which can produce
    symmetric events.  Due to the orientation of the sheet combined
    with the locations of the two crests, the lens has a symmetric column density
    profile that results in large bending angles at $x=0$.  
Note in the picture we are considering, the sheet only occupies a small percentage of the distance between the source and the observer, and the angle between
the sheet and the line-of-sight to the pulsar is much less than one.}\label{fig:ESE}
\end{figure*}

Although we have focused on pulsar scintillation, the picture
discussed within this paper can also lead to 
symmetric lensing events.  Consider two consecutive crests which
extend from the plane of the 
sheet in opposite directions.  If the crests are very close together, as
shown in Fig. \ref{fig:ESE}, then a
very high column density through the lens is achieved at $x=0$,
and the column density profile is symmetric in $x$.  Further
quantitative analysis, outside 
of the scope of this paper, is necessary to confirm the qualitative
similarities between this set-up and the observed characteristics of
symmetric lensing events such as ESEs.
In particular, the ability of this model to predict both the
light curves and frequency evolution of these events must be studied.  While these events are expected to be generic, as with any corrugated sheet there is an inclination angle for which the line of sight to the pulsar is parallel to the tangent through the sheet at $x=0$, an approximation of the prevalence of this event requires assumptions of both the distribution of plasma sheets in the ISM and the distribution of wavelengths and amplitudes of waves along those sheets, and is thus beyond the scope of the paper.  However, this
feature of the corrugated sheet model presented here may also explain why both
symmetric and and asymmetric features are seen in the echoes of
 the Crab pulsar \citep{lyne_pulsar_2001}.

\section{Conclusions}
\label{sec:conclusions}

We have investigated the effects of a corrugated plasma
sheet as the source of pulsar scintillation arcs.  Using geometric optics, we
calculate the number of images and magnifications of these images as a
pulsar moves behind a wave crest on one of these sheets.  We find that in the limit
$\theta \ll r/2$ the lens can be described by a
single parameter, $\Delta n \, \theta_T \sqrt{r}$, which can be
constrained from observations of the pulsar.  Once this parameter is known,
this model can be used to predict how the magnifications and locations of the
images will change over time and frequency, providing a concrete test
of this model.  In particular, we see that over a small band of 32 MHz
we expect changes in the locations of the images comparable to those
measured by \cite{brisken_100_2010}.  We also 
see that we expect very different behaviour from an overdense sheet or
an underdense sheet:  In the overdense case, the angular separation
between two images increases at higher frequencies, while in the
underdense case it decreases at higher frequencies.
\cite{brisken_100_2010} find that the separation decreases with
frequency for two of three identified groups of images in the
secondary spectrum of PSR B0834+06, suggesting that the lens is
underdense.  There are other qualitative differences between the
underdense and overdense lenses: magnifications greater than 1 occur only when the sheet is
overdense, both images of the pulsar disappear when the pulsar is just
behind the crest of the wave in the overdense case, and only the
overdense lens produces inverted images.  

Qualitatively, there are two major differences between this model and
previous models of ESEs and pulsar scintillation arcs.
Other 
authors
\citep[\textit{e.g.}][]{bannister_real-time_2016,clegg_gaussian_1998,pen_refractive_2012,tuntsov_dynamic_2015}
have considered smoothly varying electron column densities, which
adhere to the odd image theorem and produce one or three images.
In this work, we see that when we consider an abrupt change in density at the sheet boundary,
the column density considered is not continuously differentiable, and the lens
produces one or two images.  When we consider a smooth density profile within the sheet, our lens does adhere to the odd-image theorem.  However, due to the thinness of the sheet considered, the additional images formed are faint at angular separations of interest to pulsar scintillation studies, and can be safely ignored.
In
addition, we have considered an asymmetric lens that may
explain
asymmetric dispersion and scattering events, such as the anomalous
dispersion measure variations seen in PSR J1713+0747 
\citep{jones_nanograv_2017,lentati_from_2016} and asymmetric echo events observed in
the Crab pulsar \citep{lyne_pulsar_2001}.  

In future work, we will apply this test to
observations of pulsar scintillation and of pulsar echoes.  The simplest test comes from
observations of pulsars with well-defined inverted arclets in their
secondary spectra, as these arclets can be mapped to
images on the sky through VLBI observations \citep{brisken_100_2010}
and therefore to individual wave crests in the current sheet.  Ideal
observations will be at low frequencies where scintillation effects
are strongest and over a wide bandwidth in order to measure the
changes in the secondary spectrum over frequency.  We can also compare this model to observed changes in the
secondary spectrum over time.  For this, we desire many observations
of the pulsar on week to month long timescales.  Finally, we can
simulate and test features of this model that arise when we consider
many images of a single pulsar being created by multiple crests, such
as the scattering tail of a highly scattered pulsar.  The results of
these tests will assess the consistency of the corrugated, closely
aligned, sheet model with observations of pulsar scintillation. 

\section*{Acknowledgements}

We thank Marten van Kerkwijk and Robert Main for many valuable
discussions from the early stages of this work.  
We also thank Dan Stinebring and Barney Rickett for useful
suggestions, and Peter Martin for helpful discussions.  We thank the referee for 
comments that have much improved the paper.
DS acknowledges funding from NSERC. The Dunlap
Institute for Astronomy and Astrophysics is funded through an
endowment established by the David Dunlap family and the University of
Toronto.



\bibliographystyle{mnras}
\bibliography{paper}





\bsp 
\label{lastpage}
\end{document}